\documentclass[useAMS,usenatbib]{mn2e_mod}
\usepackage{graphicx}
\usepackage{amsmath}
\usepackage{amssymb}
\usepackage{longtable}
\usepackage{deluxetable}
\usepackage[font=small,labelfont=bf]{caption}
\usepackage{tablefootnote}
\newcommand\aj{{AJ}}%
\newcommand\araa{{ARA\&A}}%
\newcommand\apj{{ApJ}}%
\newcommand\apjl{{ApJ}}%
\newcommand\apjs{{ApJS}}%
\newcommand\aap{{A\&A}}%
\newcommand\aaps{{A\&AS}}%
\newcommand\mnras{{MNRAS}}%
\newcommand\nat{{Nature}}%
\newcommand\planss{{Planet.~Space~Sci.}}%
\title [Dusting off DIBs]{Dusting off the diffuse interstellar bands: DIBs and dust in extragalactic SDSS spectra}

\author[Baron et al.]
{Dalya Baron$^{1}$\thanks{dalyabaron@mail.tau.ac.il},
Dovi Poznanski$^{1}$\thanks{dovi@tau.ac.il},
Darach Watson$^{2}$,
Yushu Yao$^{3}$,\newauthor
\& J. Xavier Prochaska$^{4}$\\
$^{1}$School of Physics and Astronomy, Tel-Aviv University, Tel Aviv 69978, Israel.\\
$^{2}$Dark Cosmology Centre, Niels Bohr Institute, University of Copenhagen, Juliane Maries Vej 30, DK-2100 Copenhagen, Denmark.\\
$^{3}$Lawrence Berkeley National Lab, 1 Cyclotron Road, Berkeley CA 94720, USA.\\
$^{4}$Department of Astronomy and Astrophysics \& UCO/Lick Observatory, University of California, Santa Cruz, CA 95064, USA.}

\begin{document}

\maketitle

\label{firstpage}
\begin{abstract}

Using over a million and a half extragalactic spectra we study the properties of the mysterious Diffuse Interstellar Bands (DIBs) in the Milky Way. These data provide us with an unprecedented sampling of the skies at high Galactic-latitude and low dust-column-density. We present our method, study the correlation of the equivalent width of 8 DIBs with dust extinction and with a few atomic species, and the distribution of four DIBs -- $5780.6$\,\AA, $5797.1$\,\AA, $6204.3$\,\AA, and $6613.6$\,\AA\ -- over nearly $15\,000$\,deg$^2$. As previously found, DIBs strengths correlate with extinction and therefore inevitably with each other. However, we show that DIBs can exist even in dust free areas. Furthermore, we find that the  DIBs correlation with dust varies significantly over the sky. DIB  under- or over-densities, relative to the expectation from dust, are often spread over hundreds of square degrees. These patches are different for the four DIBs, showing that they are unlikely to originate from the same carrier, as previously suggested.

\end{abstract}


\vspace{1cm}
\section{Introduction}\label{s:intro}

The diffuse interstellar bands (DIBs) -- unidentified absorption features that have been detected so far in the optical and near infrared wavelength range -- are a long standing mystery in astronomical spectroscopy (e.g, \citealt{herbig95} and \citealt{sarre06}). The first DIBs were discovered in stellar spectra in 1919 \citep{heger22} and were stationary in the spectra of binary stars, thus associated with the interstellar medium \citep{merrill36, merrill38}.

So far, hundreds of DIBs have been found, and large molecules and their ions as the most likely candidate carriers (see \citealt{herbig95} for a review).  Many carriers of DIBs have been proposed since their discovery, from polycyclic aromatic hydrocarbons (PAHs) \citep{Salama99,Zhou06,Leidlmair11}, to fullerenes \citep{kroto85}, and various hydrocarbon molecules and ions such as HC$_5$N$^+$, HC$_4$H$^+$, and linear C$_3$H$_2$ (\citealt{Motylewski00}, \citealt{Kreowski10} and \citealt{Maier11} respectively). However the small number of wavelength matches that have been observed and the large number of features in interstellar spectra have so far prevented reliable identifications. Given the large number of known DIBs, a secure identification of a given carrier with a set of lines will require not only state-of-the-art laboratory spectroscopy, but also astronomical evidence that these specific lines are co-located and their strength correlated. 

DIBs studies are generally based on stellar spectra, typically at low galactic latitude and relatively high extinction (though several studies used extragalactic sources, such as stars in nearby galaxies or supernovae, to study the DIBs in the Milky-way (MW), e.g.,  \citealt{welty06,cox06b,cordiner08,Cordiner11,van-loon13}). Moreover, many studies are based on exquisite spectra, albeit only through a handful of sight lines, and thus suffer from a difficulty to draw general overarching conclusions. Recently, studies have raised these numbers to hundreds of stars (e.g., \citealt{friedman11}; hereafter F11) and even larger samples (\citealt{kos13} who studied the 8620\,\AA\ DIB using 500\,000 spectra). \citet{zasowski14} use $60\,000$ infrared spectra, \citet{yuan12} use $2000$ stellar spectra from the Sloan Digital Sky Survey (SDSS), and \citet{maiz-apellaniz14} are compiling a sample that will consist of thousands of stellar spectra.

The public domain already includes a large number of spectra that could be used to study DIBs in a statistical manner. \citet{poznanski12} (hereafter P12)  used a million SDSS spectra of galaxies and quasars to study the mean properties of the Milky Way Na\,I\,D absorption doublet. They showed that even though every spectrum has low signal-to-noise (S/N), by binning the spectra in large numbers (typically more than 3000) one can detect the doublet and measure its equivalent width (EW). As we show below, updating the data and optimizing the method of P12, we can recover and study many DIBs, through many sight lines, even though the DIBs are much weaker than the sodium doublet. 

Since our sample consists of more than 1.5 million spectra that span over $15\,000$\,deg$^2$ we are less prone to the systematics and variance that arise when one compares individual sight lines, though a different set of systematics is unavoidable. Furthermore, we are probing a high Galactic-latitude environment that has been barely studied before.

In Section \ref{s:data} and \ref{s:1} we present the data and our method, which includes  extensive simulations for uncertainty estimation. We use these in Section \ref{s:DIB-ISM} to study the relation of DIB strength with dust, sodium, and calcium, integrated over the entire SDSS footprint. in Section \ref{s:sky} we study the distribution over the sky of 4 DIBs after we correct for the general dependance on dust column density. We review and discuss our findings in \ref{s:conc}\footnote{While performing this analysis we became aware of a similar effort by \citet{lan14a}. We decided to finish the two analyses independently and submit the two papers simultaneously. Section \ref{s:conc} briefly compares their work to ours.}. All the tables, as well as the data that went into our figures can be downloaded at \texttt{www.astro.tau.ac.il/$\sim$dovip/dibs.zip}.

\begin{figure}
\includegraphics[width=3.25in]{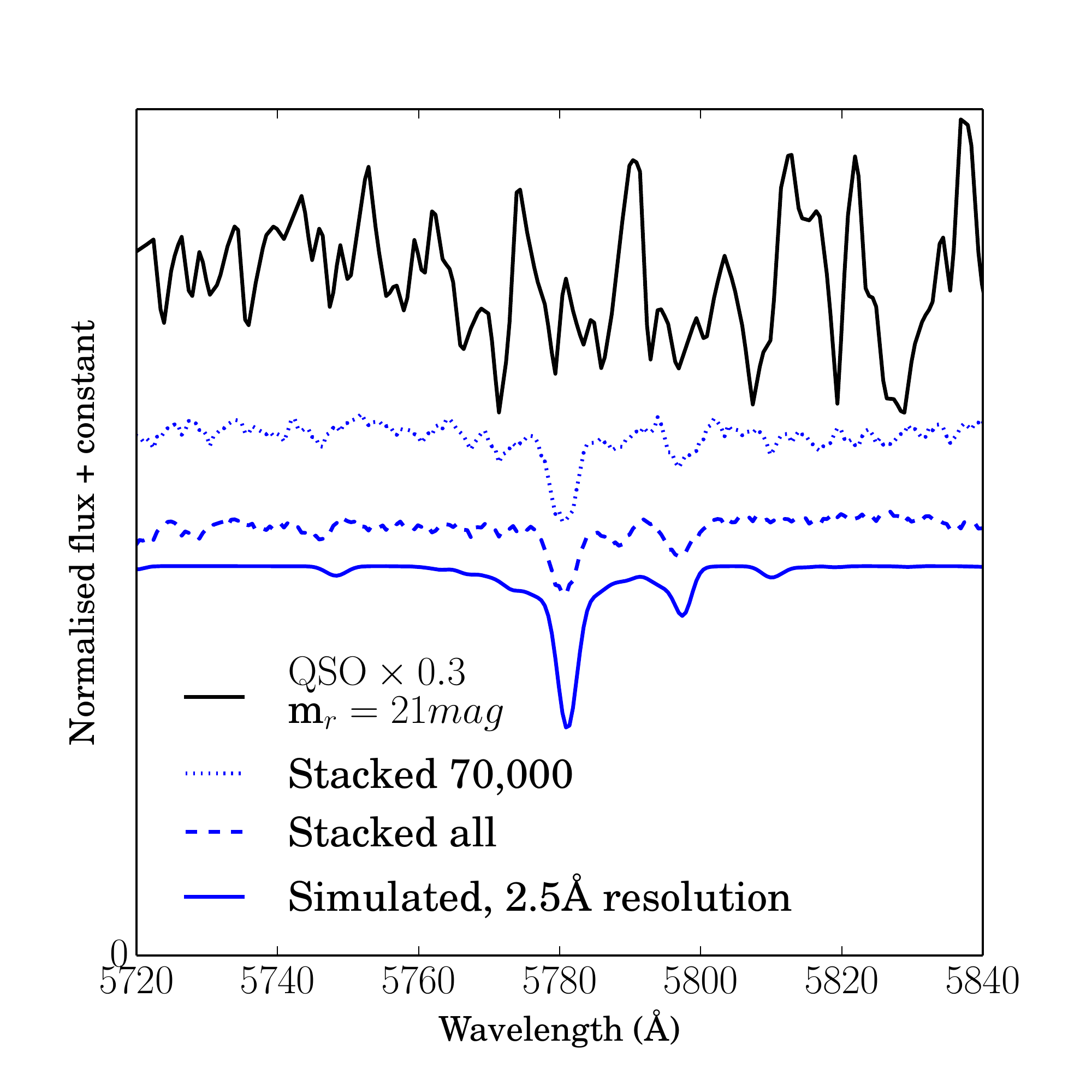}
\caption{Top to bottom: a random SDSS QSO spectrum with $m_r=21$\,mag,  multiplied by 0.3 (black), a stack of $70\,000$ spectra with a median color excess of $\mathrm{E}(B-V)=0.05$\,mag (blue dotted), a stack of the entire SDSS sample (blue dashed), and a simulated DIB spectrum (blue).}\label{f:20}
\end{figure}

\begin{figure}
\includegraphics[width=3.25in]{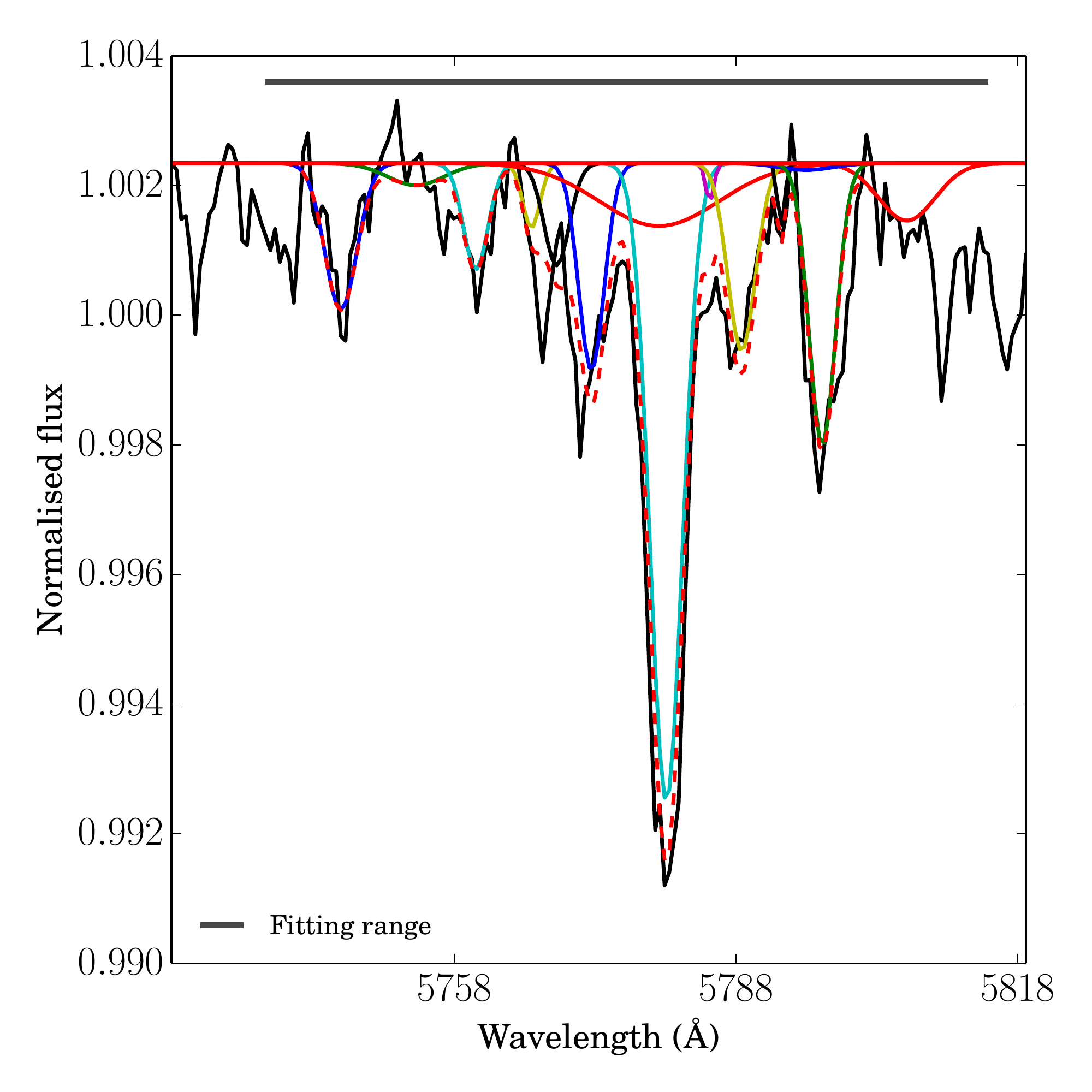}
\caption{Example fit for the 5780.6\,\AA\ DIB. Black line is the stacked spectrum, coloured lines are the individual Gaussian components. The horizontal grey line represents the limits of the fitting range as defined in section \ref{s:1}, and the red dashed line is our fit: a sum of all the Gaussians.}\label{f:21} 
\end{figure}

\begin{figure}
\includegraphics[width=3.25in]{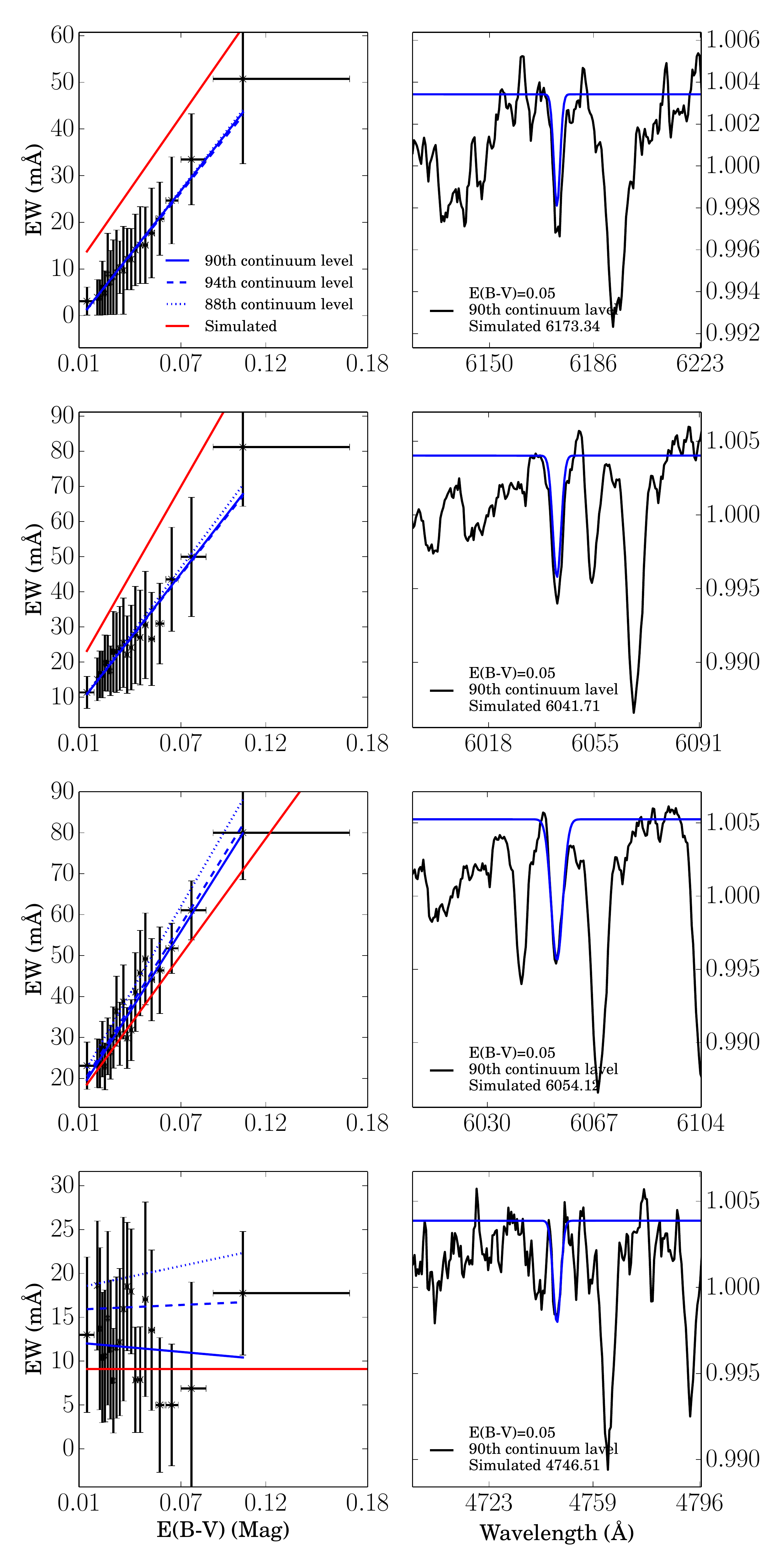}
\caption{Right: 4 representative examples of synthetic lines (out of the 10\,000 we simulate) in a typical color-excess bin (black), and best fitting Gaussian function (blue). Left: the EW of the lines vs. $\mathrm{E}(B-V)$. The black data points are the measured EW, blue lines are the best linear fit for different continuum levels (0.88, 0.9 and 0.94) while the red line is the simulated relation. The 0.9 continuum level typically works best.}\label{f:1}
\end{figure}

\section{stacking spectra}\label{s:data}

The wavelengths and basic parameters of DIBs are gathered from the catalog of \citet{jenniskens94}, later updated with new discoveries from \citet{jenniskens96} and \citet{krewski95}. We cross match these DIBs with the more recent catalog of \citet{hobbs08} and \citet{hobbs09}, discarding all lines which are not present in the more recent catalog. We remain with 175 DIBs in total.

The SDSS ninth data release \citep{ahn12} includes more than 200,000 Quasars (QSOs) and about 1.3 million galaxies with low resolution ($\frac{\lambda}{\Delta\lambda} \cong 1800$) and low S/N (4--20 per resolution element) compared to spectra typically used for DIBs studies. From these we choose spectra with redshift $z > 0.005$ in order to avoid contamination from absorption lines in the rest frame of the QSOs and galaxies. This limit translates to a minimum offset of 30\,\AA\, between host galaxy and Milky Way (MW) lines.
After excluding stellar spectra that were mistakenly identified as QSOs and galaxies by the SDSS pipeline (we identified these stars via their Balmer absorption lines), we remain with 1,307,656 galaxies and 213,959 QSOs, out to $z \approx 5$ , more than a million and a half spectra in total, about 50 percent more than P12 who used DR8. The large redshift span allows us to smear out any rest frame features. 

Although the SDSS pipeline includes correction for telluric lines, residual telluric lines may appear when binning thousands of spectra. We use TelFit \footnote{TelFit is a python wrapper to LBLRTM, the Line-By-Line Radiative Transfer Model. \texttt{http://www.as.utexas.edu/$\sim$kgulliks/projects.html}} \citep{gullikson14} to fit telluric lines and remove them. We find a non negligible telluric line residual near the $6281.1$\,\AA\, DIB and the near Na\,I\,D doublet, the remaining telluric lines we detect do not affect our fitting process. We therefore discard the $6281.1$\,\AA\, DIB from further analysis and subtract the telluric line near the Na\,I\,D doublet. 

For correlation with the column density of dust we use the maps of \citet{schlegel98} (hereafter SFD) who used all-sky measurements of infrared emission as a tracer of dust content.

We expand the method used by P12 to the entire wavelength range of the SDSS spectra: first we linearly interpolate every spectrum to an identical wavelength grid of \,$0.5$\,\AA\ resolution, in the \,$3817-9206$\,\AA\ range. We use the Savitzky-Golay (SG) smoothing algorithm \citep{savitzky64}\footnote{SG is a generalization of `running mean' where instead of fitting a constant in the window (the mean) one can fit a polynomial.} with a 2nd order polynomial fit and a moving window size of $150$\,\AA\ to fit and divide out the continuum, thus obtaining normalized fluxes. This removes all intrinsic features except for narrow lines. 
With a $150$\,\AA\ window, the smoothing removes only features that are broader than $100$\,\AA, thus even the broadest DIBs should be left untouched. We then group the spectra by their color excess value, $\mathrm{E}(B-V)$ (or coordinates in Section \ref{s:sky}), using the maps of SFD, and combine them by calculating the median at every wavelength. Calculating medians requires one to sort the fluxes, a computationally expensive procedure, but we tried several different methods of averaging with sigma clipping and got poorer S/N than with medians. 

The grouping and stacking of spectra is performed within a database we built for that purpose, using the SciDB open-source scalable array-database\footnote{\texttt{www.scidb.org}}, that allows large datasets to be accessed rapidly by parallel processes, with native complex analytics. Technical details will be published in Yao et al. (in preparation). 

In figure \ref{f:20} we show the result of stacking 70,000 spectra, as well as  stacking the entire SDSS, compared to a single spectrum. The gain in S/N is obvious. When comparing the stacked spectra to a simulated spectrum based on the DIB catalog convolved with the SDSS resolution, one can see that we recover not only the strong 5780.6\,\AA\ line, but also the much weaker line 5797.1\,\AA. The median color-excess value of the 70,000 stacked spectra (0.05) is higher than for the entire SDSS sample (0.03). 5780.6\,\AA\ is shallower in the deeper stack because its strength correlates with extinction.

\section{EW measurements and uncertainties}\label{s:1}

Since our resolution is too low to resolve any line structure we can simply fit Gaussians to measure the EW of the DIBs. Due to the large number of DIBs in the wavelength range and the low resolution of the SDSS spectra, most of the lines are blended. As a result we are limited in the number of DIBs we can measure well, a question we explore extensively via simulations later on in this section.
For each line we set the initial fitting range to be 10 times the line's full width half maximum (FWHM). In case there are no other lines (DIBs or known molecular and atomic lines from a list kindly provided by N.L.J. Cox, private communication) in the initial fitting range we define the line as isolated and the fit is straightforward.
Otherwise, for every additional line in the initial fitting range, we increase the range until it includes all the blended lines around the initial line. We then fit multiple simultaneous Gaussian functions, one for each line in the range. For every included DIB we allow the central wavelength and width to vary within the cataloged uncertainty.

One can see the process in figure \ref{f:21}: the initial DIB is 5780.6\,\AA. Since the DIB is blended we increase the fitting limits until all the blended lines are included. We then fit simultaneously 17 Gaussian functions and extract the Gaussian that represents the line of interest. We compute the EW by integrating over the best fitting Gaussian function.

To estimate our uncertainty and look for possible biases we measure the flux and the EW of synthetic lines. We create synthetic absorption lines by randomly assigning their depth, FWHM and central wavelength. The synthetic lines are Gaussian shaped with amplitudes and FWHMs that are similar to the DIBs typical values: amplitudes near $10^{-3}$ and FWHM of the order of a few \AA, for half of the Gaussians we set the amplitude to be extinction dependent. 

Accuracy should compel us to simulate the effect of our entire pipeline on the synthetic lines, which entails their introduction into the spectra before stacking. We multiply every single, original, spectrum by a synthetic spectrum that contains the synthetic lines. We then process the spectra as before, normalizing, removing broad source signal, and stacking. We compared the spectra thus obtained to the result of injecting the lines on the already-stacked spectra, and find that the resulting spectra and lines are virtually identical. We therefore simulate the lines on the stacked spectra. This requires much less computational resources (since we operate on tens of spectra, rather than hundreds of thousands), and shows that we do not introduce a significant bias during the stacking process. To probe the entire wavelength range we inject 50 lines with various atributes and measure their EWs. We repeat the process 200 times, each time randomizing the amplitude, its dependance on extinction, FWHM and central wavelength. This results in $10\,000$ synthetic absorption lines in total that map the relevant parameter space well.

Figure \ref{f:1} shows on the right four representative synthetic lines. The figure shows only the line's best fit Gaussian though all the absorption lines in the range of the plot are fitted. Clearly the synthetic lines can be detected, and we find that their EW can be measured down to 5m\AA. Around this level we consistently find that the error in EW is comparable to its value. We subsequently use this as our detection limit.

The synthetic lines allow us to measure the errors caused by uncertain continuum estimation and line blending that are typically the main causes for uncertainty underestimation (e.g., \citealt{herbig95} and F11). For example, McCall et al. (2010) speculated that the nearly perfect correlation between the DIB pair 6196 \AA\ and 6613 \AA\ can deviate from a perfect correlation, if the DIBs are indeed vibronic transitions of the same molecule, due to a systematic error caused by the continuum uncertainty. 

The SG algorithm we use to normalize the initial spectra removes very broad and low frequency features (typically broader than 100\,\AA). Decreasing the window size of the SG algorithm results in removal of narrower features, which we want to avoid in order to keep the absorption lines intact. We therefore do not get a continuum level that is perfectly equal to 1. We account for these offsets by fitting for the continuum level. We try the following continuum-level definitions: fitting a linear and 2nd polynomial order curve to the fitting range, setting the continuum level to be a constant at the 88th-98th percentile of the flux in the fitting range, or forcing the continuum level to be unity in every fitting range. For every method we measure the EW of the synthetic lines and compare it to the EW we introduced. One can see in figure \ref{f:1} the EW of the synthetic lines vs. $\mathrm{E}(B-V)$ for the 88th, 90th and 94th percentile continuum level. We find that defining the continuum level to be constant as the 90th percentile of the flux in the fitting range predicts the initial EW with the smallest scatter. Furthermore, we find that for that same definition there is no systematic offset in the fluxes we measure. We therefore use it, as it minimizes both statistical and systematic errors. 

\begin{figure}
\includegraphics[width=3.2in]{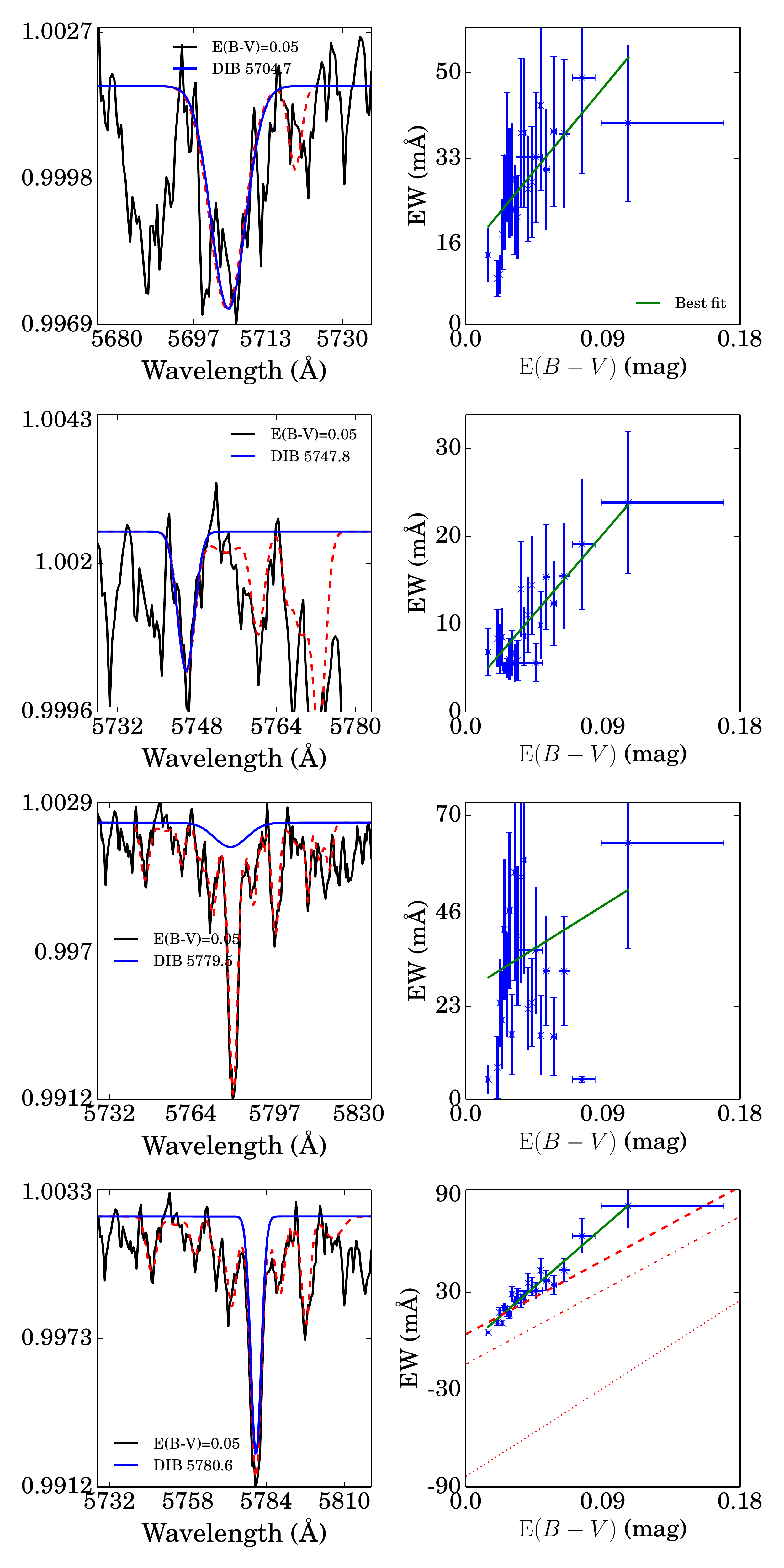}
\caption{Left panel - median spectrum at a typical E(B-V) value (black), the Gaussian fit for the single DIB (blue), and full fit including other components (red dashed). Right panel - measured EW vs. $\mathrm{E}(B-V)$ (blue data points), the best linear fit we deduced (green solid) and the slopes F11 and K13 deduced ($\zeta$ and $\sigma$ lines of sight)  in red dashed, doted and dash-doted, respectively.
When the relation is clearly is not linear, we restrict our fit to the linear range.}\label{f:2}
\end{figure}

\begin{figure}
\includegraphics[width=3.2in]{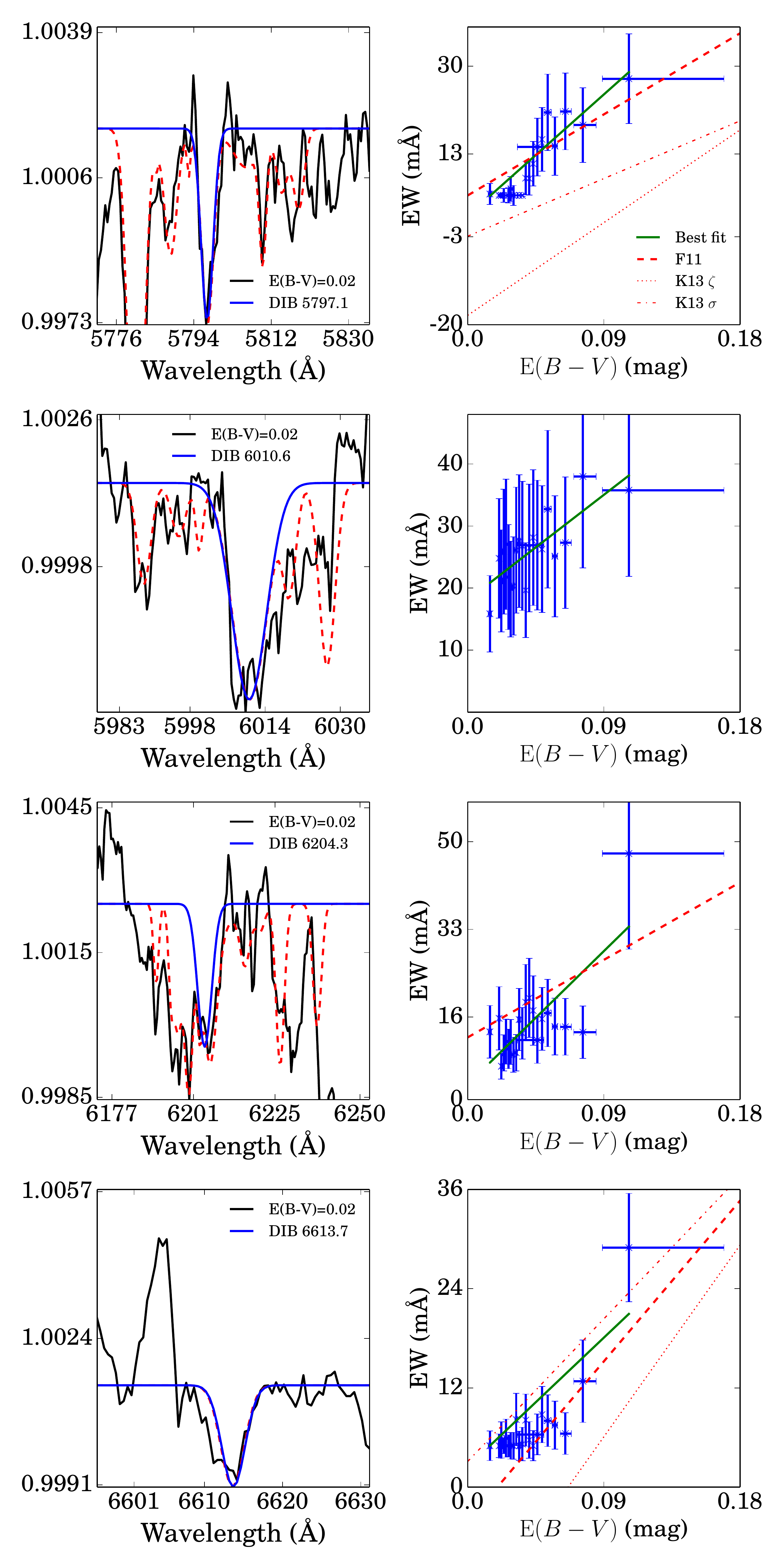}
\caption{Same as Figure \ref{f:2} for 4 additional DIBs.}\label{f:added}
\end{figure}

P12 found a systematic shift, where their EW measurements on the SDSS spectra were consistently underestimated by about 10 percent. They assigned this shift to a bias that arises from the uneven distribution of the sample which is skewed towards low extinction spectra. Although we do not find a systematic shift in the DIBs central depth at the 90th continuum level, we find a systematic shift of 3 percent between the measured and introduced EW, indicating that our line widths are slightly underestimated during the Gaussian fitting. To account for this systematic shift in EW measurement, which is a few times smaller than our typical statistical uncertainty, we add a 3 percent correction to every EW value derived from the stacked spectra.

Using the $10\,000$ synthetic lines, we measure the residual between the simulated and measured EW, and determine the median error as a function of the EW, the wavelength -- accounting for the effects of blending with nearby lines, and the goodness of the fit to the data. We use these as estimates of the uncertainty in the EW measurements of the real lines as a function of their own measured EW, wavelength, and goodness of fit. We find that typically, for EWs between 60 and 1000\,m\AA, our uncertainties are of order 15--20 percent. This is a non-negligible fraction, but it is due in large parts to blending, and it accounts for all the major sources of uncertainty, including the one stemming from continuum determination that is often not quantitatively taken into account in past works. At lower EWs and in many parts of the spectrum where the density of lines is significant, our EWs are  unreliable. Moreover, our simulations do not account for false positives detections, which could occur for weak lines. This limits our subsequent analysis to a few strong or relatively isolated lines. 

Since some of the DIBs we fit are weaker than our noise threshold one may argue that they should not be included in the fitting process of blended lines. We therefore compare our fitting process to fitting only DIBs that are stronger than the noise threshold. We use the lowest reddening bin (mean color-excess 0.0146) and simulate absorption lines with different strengths, and a central wavelength that corresponds to the 8 DIBs we measure in section \ref{s:DIB-ISM}. We compare the measured EW to the one we simulate, using both fitting methods, and consistently find a bias in the method that ignores weak lines, showing that our method is more precise.

\begin{table}
	\tiny
\setlength{\tabcolsep}{3pt}
\caption{Sample of DIBs EW as a function of E(B-V)} % title of Table 
%\centering % used for centering table 

\begin{tabular}{lccccc} % centered columns (4 columns) 
\hline \hline %inserts double horizontal lines 

DIB[\AA]\tablenotemark{a} & $\mathrm{E}(B-V)$ & $\Delta^{+} \mathrm{E}(B-V)$ & $\Delta^{-} \mathrm{E}(B-V)$ & EW[m\AA]& $\Delta$EW[m\AA] \\ [0.5ex]
\hline
$5780.6$ & $0.0147$ & $0.00106$ & $0.00101$ & $5.31$ & $0.88$ \\
$5780.6$ & $0.0208$ & $0.00084$ & $0.00078$ & $11.74$ & $1.94$ \\
$5780.6$ & $0.0223$ & $0.00074$ & $0.00067$ & $17.6$ & $2.91$ \\
$5780.6$ & $0.0238$ & $0.00078$ & $0.00073$ & $11.11$ & $1.83$ \\
$5780.6$ & $0.0253$ & $0.00116$ & $0.00107$ & $20.23$ & $3.34$ \\
$5780.6$ & $0.0269$ & $0.0013$ & $0.00119$ & $18.17$ & $3.0$ \\
$5780.6$ & $0.0285$ & $0.00143$ & $0.00142$ & $16.5$ & $2.72$ \\
$5780.6$ & $0.0302$ & $0.00102$ & $0.00092$ & $28.87$ & $4.77$ \\
$5780.6$ & $0.0321$ & $0.00091$ & $0.00082$ & $23.76$ & $3.92$ \\
$5780.6$ & $0.034$ & $0.00096$ & $0.00089$ & $27.86$ & $4.6$ \\
$5780.6$ & $0.0361$ & $0.0007$ & $0.00064$ & $24.81$ & $4.1$ \\
$5780.6$ & $0.0384$ & $0.00069$ & $0.00061$ & $26.73$ & $4.41$ \\
$5780.6$ & $0.0407$ & $0.0007$ & $0.00061$ & $35.76$ & $5.9$ \\
$5780.6$ & $0.0433$ & $0.00067$ & $0.00061$ & $33.59$ & $5.54$ \\
$5780.6$ & $0.046$ & $0.0132$ & $0.00427$ & $31.07$ & $5.13$ \\
$5780.6$ & $0.0491$ & $0.00068$ & $0.00062$ & $43.52$ & $7.18$ \\
$5780.6$ & $0.0528$ & $0.00238$ & $0.00253$ & $37.35$ & $6.17$ \\
$5780.6$ & $0.0576$ & $0.00171$ & $0.00175$ & $34.59$ & $5.71$ \\
$5780.6$ & $0.0646$ & $0.00341$ & $0.00389$ & $43.75$ & $7.22$ \\
$5780.6$ & $0.0762$ & $0.00607$ & $0.00868$ & $64.77$ & $10.69$ \\
$5780.6$ & $0.1065$ & $0.01754$ & $0.06285$ & $83.32$ & $13.75$ \\

\hline %inserts single line 
\hspace{-0.1cm}
\tablenotetext{a}{Full data available in the online version.}
\end{tabular} 
\label{t:EW_mes_small}
\end{table}

\begin{table}
\setlength{\tabcolsep}{3pt}
\caption{DIB-dust fit parameters} % title of Table 
%\centering % used for centering table 
\begin{tabular}{l c c c} % centered columns (4 columns) 

\hline\hline %inserts double horizontal lines 
DIB & Intercept & \emph{A}\tablenotemark{a} & \emph{B}\tablenotemark{a} \\ [0.5ex] % inserts table 
%heading 
\hline % inserts single horizontal line 

5704.7 & $1.7\sigma$ & $0.36\pm0.17$ & $0.0141\pm0.0081$ \\
5747.8 & $0\sigma$ & $0.20\pm0.16$ & $0.0021\pm0.0083$ \\
5779.5 & $6.3\sigma$ & $0.235\pm0.089$ & $0.0266\pm0.0042$ \\
5780.6 & $1\sigma$ & $0.813\pm0.073$ & $-0.0032\pm0.0034$ \\
5797.1 & $0.1\sigma$ & $0.261\pm0.090$ & $0.001\pm0.011$ \\
6010.6 & $2.2\sigma$ & $0.18\pm0.17$ &  $0.0181\pm0.0081$ \\
6204.3 & $0.4\sigma$ & $0.28\pm0.17$ & $0.0030\pm0.0081$ \\
6613.7 & $0.25\sigma$ & $0.17\pm0.15$ & $0.0025\pm0.0098$ \\

\hline %inserts single line 
\end{tabular} 

\tablenotetext{a}{Parameters for the best linear fit of DIB EW vs. $\mathrm{E}(B-V)$: $\mathrm{EW}[$\AA$] = A \cdot \mathrm{E}(B-V) + B$.}

\label{t:1}
\end{table}

\begin{figure*}
\includegraphics[width=0.9\textwidth]{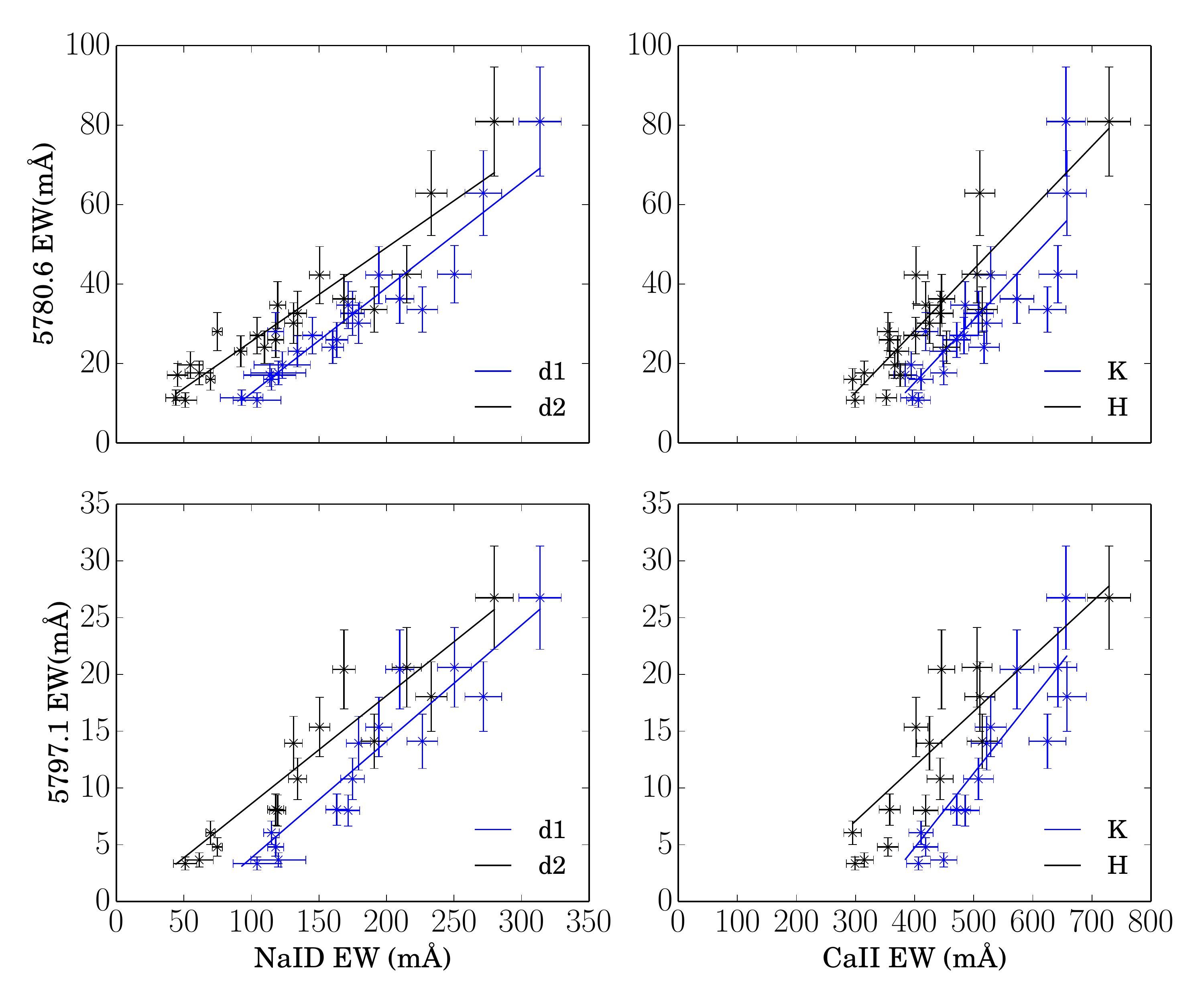}
\caption{DIB EW -- 5780.6\,\AA\ (top) and 5797.1\,\AA\ (bottom) -- versus atomic EW -- Na\,I\,D1 and D2 (left) and Ca\,II H\&K (right).  Lines are the best linear fit for the different components.}\label{f:3}
\end{figure*}

\section{Correlations of DIB\MakeLowercase{s} with other ISM species}\label{s:DIB-ISM}

\subsection{DIBs and dust extinction}\label{s:dust}

Since their discovery studies have shown that DIBs EW correlate with reddening, though with a non negligible scatter \citep{merrill38}. As the S/N increased and the measurements were refined, it became apparent that much of the scatter is intrinsic and not due to measurement errors (see \citealt{kos13a} for example; hereafter K13). 

To measure the EW as a function of dust column density we bin the spectra over similar values of $\mathrm{E}(B-V)$ as derived from the SFD maps. We tested various numbers and sizes of bins.
For $0.02 < \mathrm{E}(B-V) < 0.3 $, where we have the majority of the spectra, we fix the number of objects per bin. The number of bins for this color-excess range was determined by varying it from 10 to 40, every time comparing the resulting EW to the extinction. We found that the S/N and the results are always consistent and stable and do not depend on the number of bins. 
However, we have only 1500 spectra at $\mathrm{E}(B-V) > 0.3 $, spanning a large range of color-excesses. Since 1500 spectra are not sufficient for obtaining the necessary S/N for DIBs measurement, we exclude this range from further analysis.
All the spectra at $\mathrm{E}(B-V) < 0.02$ (nearly 400,000), where the EW is very small and difficult to measure, are stacked in a single bin.
Below, we present results using 24 bins, with 70000 spectra per bin, that seem to sample the relation well. 
We note that our last bin reaches $\mathrm{E}(B-V) = 0.3 $, but the median color excess of its spectra is close to $\mathrm{E}(B-V) = 0.11 $, so that our effective dynamic range is between no extinction and $\mathrm{E}(B-V) \sim 0.15$. 

Figures \ref{f:2} and \ref{f:added} show the EW vs. $\mathrm{E}(B-V)$ relations for the 8 DIBs that can be measured given our S/N. We only present DIBs for which the uncertainties in the linear parameters are smaller than the parameters. The 6204.3\,\AA\ and 6203.2\,\AA\ pair is blended, the EW of the 6203.2\,\AA\ DIB is below our detection threshold and we discard it from the analysis. Table \ref{t:EW_mes_small} includes a sample of the measurements (all available in the online version). Table \ref{t:1} includes the best fitting linear parameters. 

Some of the DIBs in table \ref{t:1} have an EW-dust relation that does not intercept the origin. While a negative intercept with the y-axis is often interpreted as an energy threshold effect (see F11), we usually find positive intercepts, which would indicate that some DIBs can exist in regions devoid of dust. Similarly, the $2175$\,\AA\ bump in the extinction curve of galaxies, which has often been discussed as related to DIBs, shows a decoupling from dust extinction \citep{zafar12}.

The $5780.6$\,\AA\ and $5797.1$\,\AA\ DIBs are believed to be found in diffuse clouds, with the $5780.6$\,\AA\ DIB appearing stronger where the UV radiation field is stronger and the $5797.1$\,\AA\ DIB appearing stronger in UV shielded fields, supposedly being destroyed by UV radiation, as seen in the MW and other galaxies \citep{kreowski98, cami97, cox06b, welty06}. However, \citet{Cordiner11} found a negative correlation between the strength of $5780.6$\,\AA\ and UV flux in M31, suggesting that the carrier of this DIB is destroyed by UV radiation. They note that this could instead be evidence for the production of enhanced quantities of DIBs carriers.
Studies have also suggested that DIB ratios might be linked to UV irradiation and extinction and argued that the scatter in the EW-$\mathrm{E}(B-V)$ relations of DIBs is due to different radiation fields in interstellar clouds (see \citealt{Krelowski92}, \citealt{cami97} and K13). K13 addressed this issue by dividing their spectra into the 2 known groups, $\sigma$ (not UV shielded) and $\zeta$ (UV shielded) sight lines, based on the EW ratio between $5780.6$\,\AA\ and $5797.1$\AA, claiming that intermediate sight lines must also exist since the transition between the two types should be smooth. They showed that dividing the spectra into $\sigma$ and $\zeta$ sight lines reduces the scatter in EW-reddening relation and that the behavior of the EW and reddening differs for some of the DIBs ($4964$\,\AA, $5797$\,\AA, $5850$\,\AA, $6090$\,\AA, $6379$\,\AA, $6613$\,\AA\ and $6660$\,\AA) for the different sight lines. F11 deduced EW-reddening relation with a slope that is within the range of the two types of sight lines (as also found by others, e.g. \citealt{yuan12} and \citealt{phillips13}).

Unlike the intercepts we find, K13 find negative intercepts for the $5797.1$\,\AA\ and $5780.6$\,\AA\ DIBs as shown in figures \ref{f:2} and \ref{f:added}. They note that this seems to be characteristic to all observed DIBs for the $\zeta$ sight lines due to the fact that UV shielded regions can only exist where dust column density is high enough. Had they forced the fit to pass through the origin, then their slope would have been lower. 
The slopes of the EW and reddening relation we find for the $5797.1$\,\AA\ and $5780.6$\,\AA\ DIBs above are even higher than K13 steepest $\zeta$ sight lines (the slope for $5797.1$\,\AA\: $0.199\pm0.020$  and for $5780.6$\,\AA\: $0.603\pm0.056$) by 30 percent, although only the difference for the $5780.6$\,\AA\ DIB is significant (see also \citealt{vavloon09}, who find a $\sigma$-type behavior in the Galactic Disc-Halo interface). Since the relation we obtained is based on binning different spectra and averaging them we would expect our relations to be intermediate. Furthermore, we probe an environment that is not UV shielded, with very low column densities.

\citet{mcintosh93} studied the strength of the DIBs $4428$\,\AA, $5780.6$\,\AA\ and $5797.1$\,\AA\ as a function of Galactic latitude. They found that the strength of $4428$\,\AA\ relative to the others is greatest at low latitude and decreases with increasing latitude whereas the strength of $5797.1$\,\AA\ is greatest at high latitude. \citet{van-loon13}, who use over 800 stars to study the high Galactic latitude DIBs and in the LMC, find that the growth of the EW of the $4428.1$\,\AA\, appears to slow down as the EW of the $5780.6$\,\AA\, DIB grows. These findings may explain the absence of the usually strong DIB at $4428$\,\AA\ in our spectra, which are mostly at high latitude. Furthermore, we find that $5780.6$\,\AA\ and $5797.1$\,\AA\ tend to have higher EW per reddening unit compared to the more UV shielded, Galactic plane. Our results therefore reinforce the findings of \citet{mcintosh93}.

However, a simple comparison between our sample and all the studies discussed above is misleading, since most of their sight lines are at high extinction which dominates the fit. F11 do have many stars we can compare to, and it seems clear that we tend to measure EWs which are larger than they measure for $5797.1$\,\AA\ and $5780.6$\,\AA\ DIBs. Our spectra are integrated over multiple sight lines and clouds, and low extinction, so that we mostly probe inter-cloud gas. The scale height of dust in the Milky Way is of a few hundreds pc \citep[e.g., ][]{drimmel01}, but the coronal gas (or hot ionized medium) can be found up to 10 times farther away at high Galactic latitude. If the scale height of DIBs is larger than that of dust -- if DIBs can reside further into the harsh halo, or at least somewhat outside the dusty disk -- then we expect their EW-$\mathrm{E}(B-V)$ relation to have both steeper slopes and larger intercepts in our sample than previously measured at lower Galactic latitudes, as we integrate over higher column densities of DIB carriers for a given dust column density. Recently, \citet{kos14} independently reached a similar conclusion, using pseudo three dimensional maps of the $8620$\,\AA\, DIB.

We examine whether we can find a dependence of the dust-DIB relations on Galactic latitude. We split our data into three bins of latitude, above $45\,\deg$ (56 percent of the spectra), below $-45\,\deg$ (13 percent), and in between (31 percent). Within each bin we measure the EW of the 5780.6\,\AA\ as a function of color excess. Only below $-45\,\deg$ we find a different slope than the $0.8\pm0.07$ we find over the entire sky. Instead we find a slope for the best fitting line that is even higher, $1.4\pm0.2$. This results should be taken as indicative only, since it is based on a small number of spectra, but it hints that the correlation between the dust and DIB-carrier column densities may be quite different in different regimes, even at supposedly similar extinctions. As we show in section \ref{s:sky}, the DIB-dust relation can change dramatically between sight lines, in a non-trivial way that is not necessarily correlated with latitude. 

\begin{table*}
\setlength{\tabcolsep}{3pt}
\caption{EW measurements for atoms} % title of Table 
%\centering % used for centering table 
\begin{tabular}{lcccccc} % centered columns (4 columns) 
\hline\hline %inserts double horizontal lines 

$\mathrm{E}(B-V)$ & $\Delta^{+} \mathrm{E}(B-V)$ & $\Delta^{-} \mathrm{E}(B-V)$ & NaID1 [m\AA]& NaID2 [m\AA] & CaII K [m\AA] & CaII H [m\AA] \\ [0.5ex]
\hline
0.0208 & 0.00084 & 0.00078 & $92.8\pm15.8$ & $44.0\pm7.5$ & $396.2\pm19.8$ & $352.0\pm17.6$ \\0.0223 & 0.00074 & 0.00067 & $113.6\pm19.3$ & $45.3\pm7.7$ & $384.1\pm19.2$ & $375.5\pm18.8$ \\0.0238 & 0.00078 & 0.00073 & $104.2\pm17.7$ & $51.0\pm8.7$ & $406.6\pm20.3$ & $299.3\pm15.0$ \\0.0253 & 0.00116 & 0.00107 & $122.8\pm20.9$ & $54.9\pm9.3$ & $394.2\pm19.7$ & $366.3\pm18.3$ \\0.0269 & 0.0013 & 0.00119 & $119.9\pm20.4$ & $61.4\pm10.4$ & $449.3\pm22.5$ & $314.7\pm15.7$ \\0.0285 & 0.00143 & 0.00142 & $114.8\pm5.7$ & $69.7\pm3.5$ & $410.8\pm20.5$ & $295.0\pm14.7$ \\0.0302 & 0.00102 & 0.00092 & $117.9\pm5.9$ & $74.7\pm3.7$ & $418.6\pm20.9$ & $354.8\pm17.7$ \\0.0321 & 0.00091 & 0.00082 & $134.0\pm6.7$ & $92.1\pm4.6$ & $448.0\pm22.4$ & $371.4\pm18.6$ \\0.034 & 0.00096 & 0.00089 & $145.2\pm7.3$ & $104.2\pm5.2$ & $483.5\pm24.2$ & $401.9\pm20.1$ \\0.0361 & 0.0007 & 0.00064 & $160.2\pm8.0$ & $109.7\pm5.5$ & $517.5\pm25.9$ & $454.0\pm22.7$ \\0.0384 & 0.00069 & 0.00061 & $163.2\pm8.2$ & $118.0\pm5.9$ & $471.3\pm23.6$ & $357.9\pm17.9$ \\0.0407 & 0.0007 & 0.00061 & $171.6\pm8.6$ & $119.4\pm6.0$ & $485.5\pm24.3$ & $418.9\pm20.9$ \\0.0433 & 0.00067 & 0.00061 & $174.8\pm8.7$ & $134.1\pm6.7$ & $508.1\pm25.4$ & $443.1\pm22.2$ \\0.046 & 0.0132 & 0.00427 & $179.3\pm9.0$ & $131.2\pm6.6$ & $521.9\pm26.1$ & $425.1\pm21.3$ \\0.0491 & 0.00068 & 0.00062 & $194.3\pm9.7$ & $150.5\pm7.5$ & $528.8\pm26.4$ & $402.3\pm20.1$ \\0.0528 & 0.00238 & 0.00253 & $209.9\pm10.5$ & $168.6\pm8.4$ & $573.1\pm28.7$ & $445.6\pm22.3$ \\0.0576 & 0.00171 & 0.00175 & $226.5\pm11.3$ & $190.9\pm9.5$ & $624.9\pm31.2$ & $514.5\pm25.7$ \\0.0646 & 0.00341 & 0.00389 & $250.3\pm12.5$ & $215.0\pm10.8$ & $642.4\pm32.1$ & $505.7\pm25.3$ \\0.0762 & 0.00607 & 0.00868 & $271.7\pm13.6$ & $233.2\pm11.7$ & $657.7\pm32.9$ & $510.4\pm25.5$ \\0.1065 & 0.01754 & 0.06285 & $313.7\pm15.7$ & $279.8\pm14.0$ & $656.1\pm32.8$ & $729.0\pm36.4$ \\

\\[10pt]

\hline %inserts single line 
\end{tabular} 
\\[10pt]
EW measurments as a function of E(B-V) for atoms.
\label{t:EW_mes_atoms}
\end{table*}

\begin{table}
\setlength{\tabcolsep}{3pt}
\caption{DIB-atom fit parameters} % title of Table 
%\centering % used for centering table 
\begin{tabular}{l l c c c} % centered columns (4 columns) 
\hline\hline %inserts double horizontal lines 
DIB & Atom & \emph{A} & \emph{B} & Corr \\ [0.5ex] % inserts table 
%heading 
\hline % inserts single horizontal line 

5780.6 & NaID1 & $0.265\pm0.0006 $ & $-13.962\pm0.1159 $ & -0.94\\
5780.6 & NaID2 & $0.2362\pm0.0006 $ & $1.9305\pm0.0807 $ & -0.88\\
5780.6 & CaII K & $0.1583\pm0.0004 $ & $-48.1441\pm0.2171 $ & -0.98\\
5780.6 & CaII H & $0.1549\pm0.0004 $ & $-33.7255\pm0.1695 $ & -0.97\\
5797 & NaID1 & $0.1025\pm0.001 $ & $-6.408\pm0.2043 $ & -0.97\\
5797 & NaID2 & $0.0951\pm0.0009 $ & $-0.9043\pm0.1548 $ & -0.94\\
5797 & CaII K & $0.0655\pm0.0006 $ & $-21.4761\pm0.3584 $ & -0.99\\
5797 & CaII H & $0.0482\pm0.0005 $ & $-7.391\pm0.2271 $ & -0.97\\ [1ex]

 % [1ex] adds vertical space  
 
\hline %inserts single line 
\end{tabular} 
\\[10pt]
Parameters for the best linear fit of atomic line versus DIB EWs: $\mathrm{EW(DIB)}[$\AA$] = A \cdot \mathrm{EW(\mathrm{atom})}[$\AA$] + B$ and the correlation coefficient.
\label{t:2}
\end{table}

\subsection{DIBs vs. NaID, CaII}

By construction our spectra include only contributions from the ISM without a stellar component. This allows us to derive more robust relations between DIBs and atomic species. We measure the EWs of interstellar Na\,I\,D doublet and Ca\,II H\&K lines and plot them against the EWs of the strongest DIBs: 5780.6 \AA\ and 5797.1\AA. We present the plots in figure \ref{f:3}, the EW measurements in table \ref{t:EW_mes_atoms} and the parameters of the best linear fit in table \ref{t:2}. These relations should only be applied in the linear regime of the curve of growth which we probe here,  at E$(B-V) < 0.15$. 

Clearly there is a significant correlation between all these constituents, as found before \citep[e.g.,][]{herbig93,kos13a}. This is to be expected if the DIBs and known atomic species are co-located.

\section{Mapping DIB\MakeLowercase{s}}\label{s:sky}
\subsection{Tiling the sky}
\begin{figure*}
\includegraphics[width=0.9\textwidth]{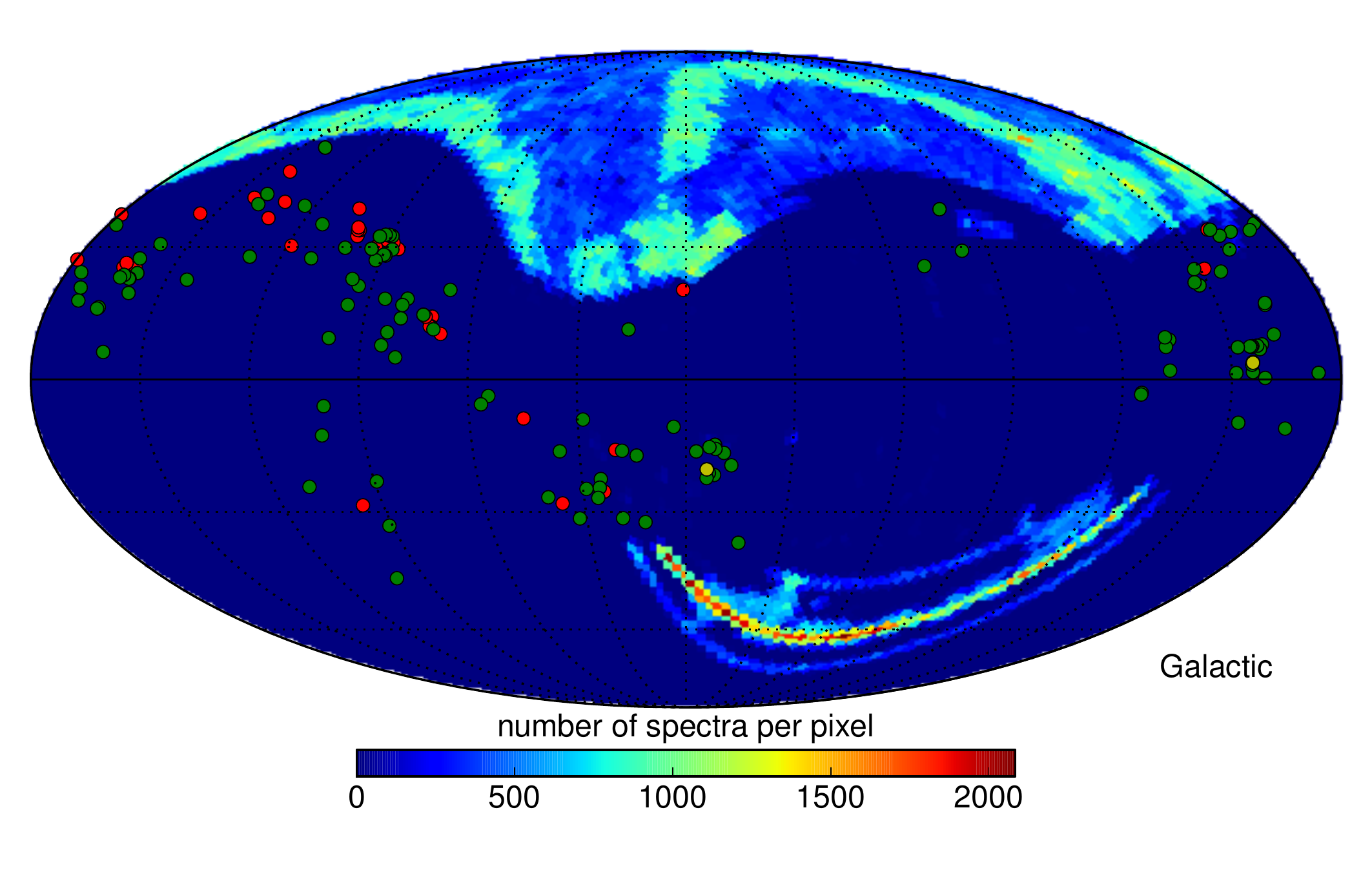}
\includegraphics[width=0.9\textwidth]{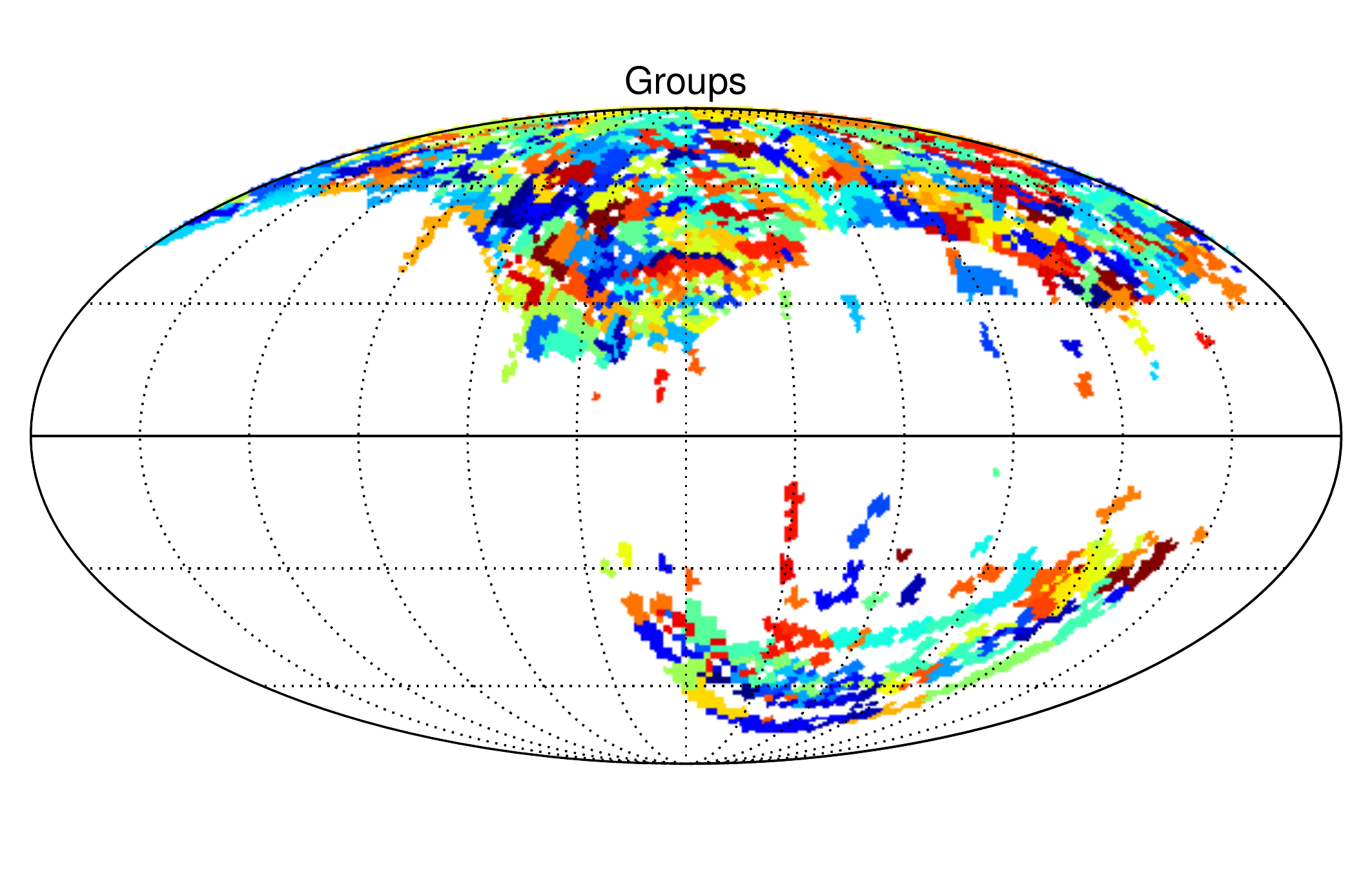}
\caption{Top: number of spectra per 3.35 deg$^2$ pixel. The red and green circles represent the coordinates of the stars used by K13 and F11 respectively. Bottom: sky map after grouping in order to reach sufficient S/N. Groups have a typical area of 20--40\,deg$^2$.}\label{f:7}
\end{figure*}

Figure \ref{f:7} shows the distribution of our sample in Galactic coordinates, compared to the samples of F11 and K13. Clearly, we sample much higher Galactic latitudes and span a much larger, and continuous, area on the sky. The average density of spectra in SDSS is about 100 spectra per deg$^2$. In order to obtain a sufficiently high S/N for detecting DIBs we must stack hundreds or thousands spectra, even when restricting ourselves to the few strongest lines. However, in order to study the distribution of the DIBs as a function of sky coordinates with the highest possible resolution we wish to group as few sight lines as possible.

We optimize as follows. We use healpix\footnote{http://healpix.sourceforge.net/} that provides an algorithm for subdividing the surface of a sphere into equal-area pixels which are arranged into lines of equal latitude \citep{gorski05}. We start by dividing the sky into $12\,288$ Healpix pixels, each pixel with an area of 3.35 deg$^2$.  We group the pixels using a modified closest-neighbor algorithm using extinction from SFD as a proxy for similarity. Every pixel starts as a one sized group and the group's extinction is defined as the average extinction over all the spectra in it. For every group that contains less than 5000 spectra, we select the closest extinction group out of the nearest neighbors, merging the groups to a single group and updating the mean extinction. We continue merging groups until every group contains a sufficient number of spectra. The grouping procedure depends on the pixel ordering, i.e. different pixel ordering could yield different groups. We adopt the healpix ring ordering system (see \citealt{gorski05} for details).

The bottom part of figure \ref{f:7} shows the final 250 groups we obtain. Most of the groups cover an area of 20--40\,deg$^2$ and a small number groups (roughly 30 groups) cover an area of 50--100\,deg$^2$.
In figure \ref{f:9} we present the average color excess and its standard deviation per group. One can see that the majority of the groups have a relatively small scatter, typically smaller than 20 percent.

\subsection{The distribution of DIBs on the sky}

From here on we study the following DIBs: 5780.6, 5797.1, 6204.3 and 6613.7 \AA, which are all well-studied, strong, and isolated enough to be detected with a relatively small number of spectra being stacked. The stacking and EW measurement procedure are identical to the one described in Section \ref{s:data}. Since we use bins with a somewhat smaller number of spectra we find that simulated DIBs are detectable when their EW is higher than 7\,m\AA, rather than the 5\,m\AA\ found before. 

\begin{figure*}
\includegraphics[width=0.9\textwidth]{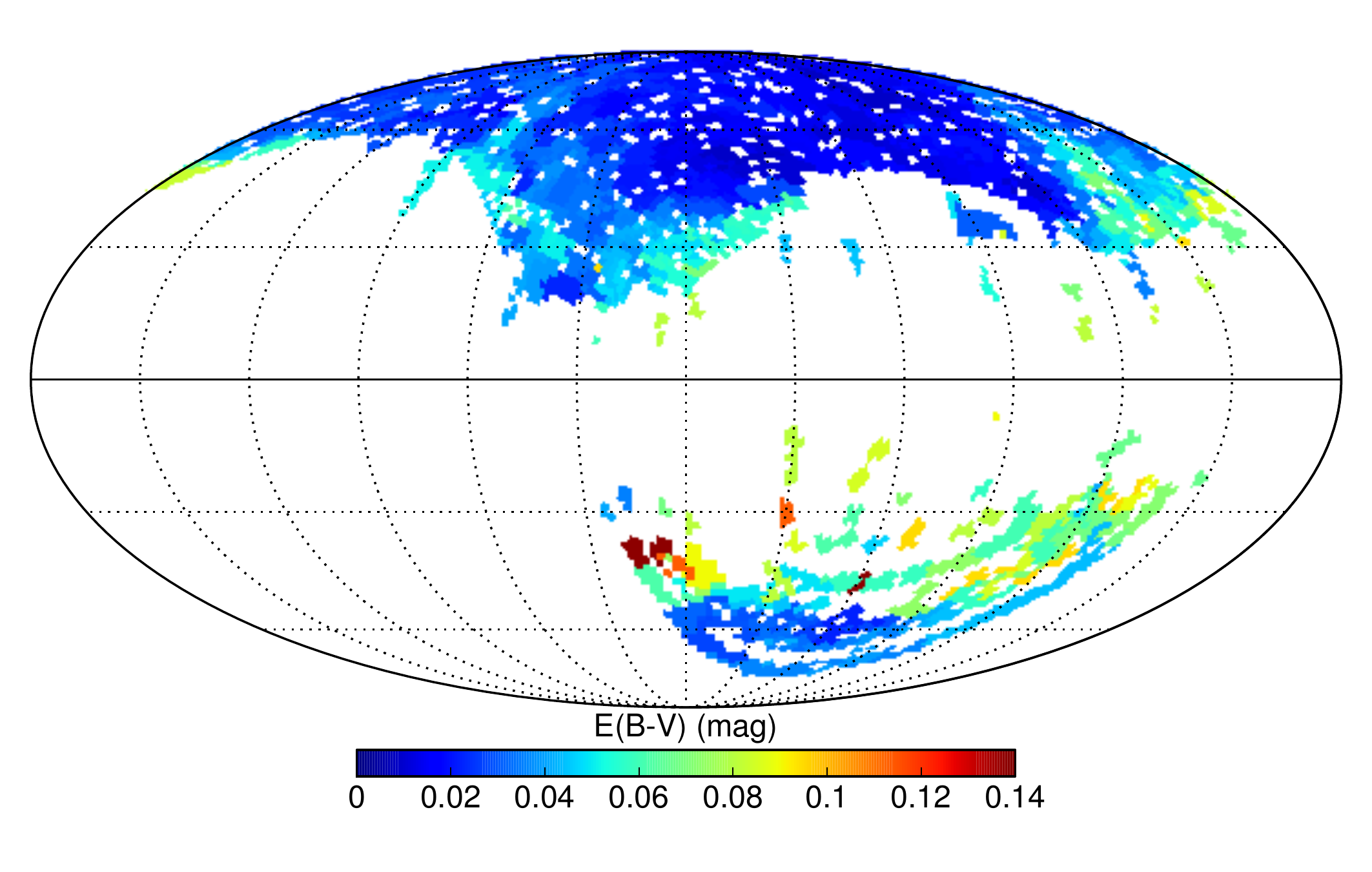}
\includegraphics[width=0.9\textwidth]{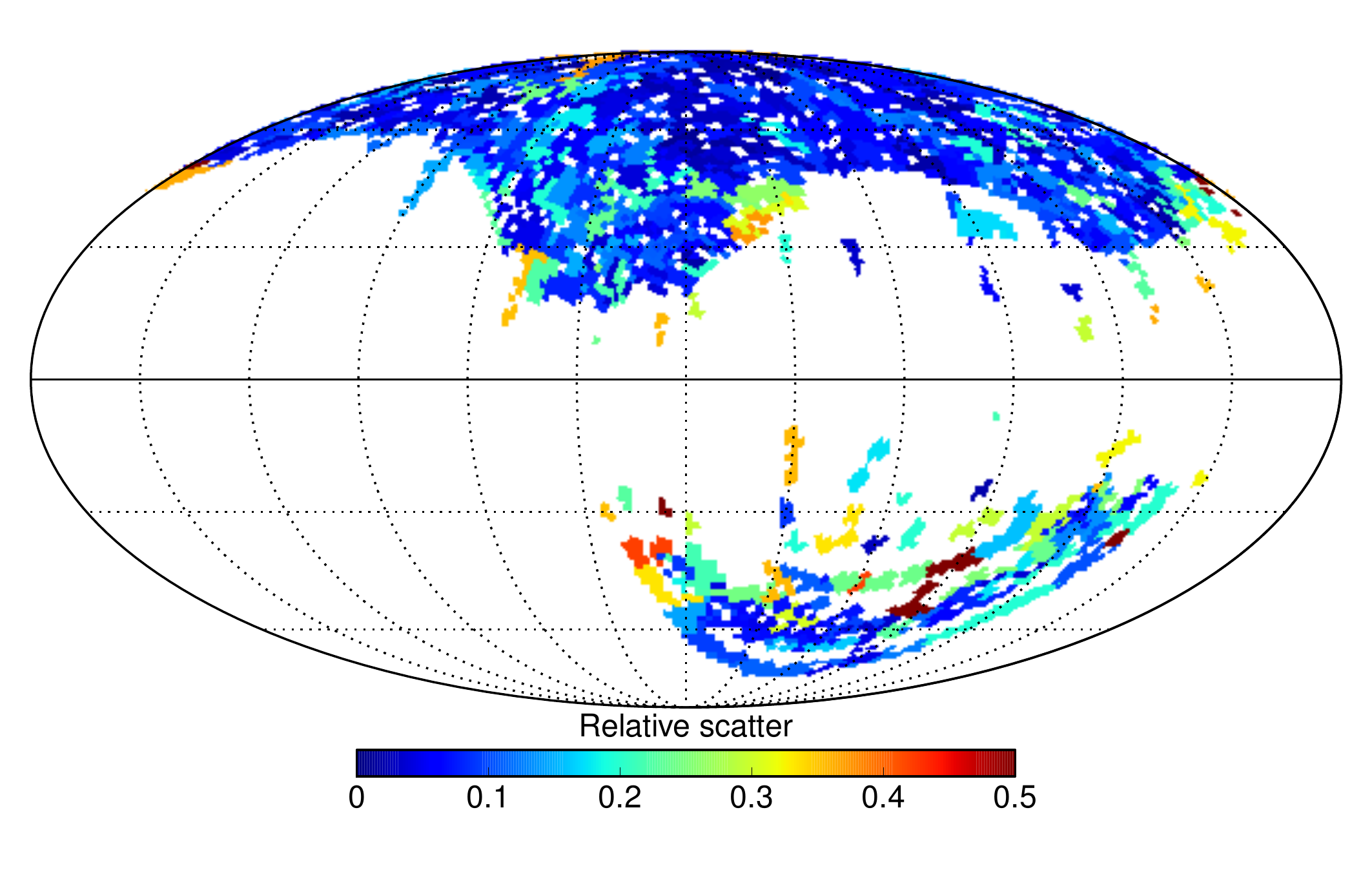}
\caption{Average color excess of every group (top), and relative scatter in each group (bottom; $\sigma/\mathrm{mean}$). The majority of groups are rather uniform, with a color excess that varies by less than 20 percent.}\label{f:9}
\end{figure*}

\begin{figure*}
\includegraphics[width=0.9\textwidth]{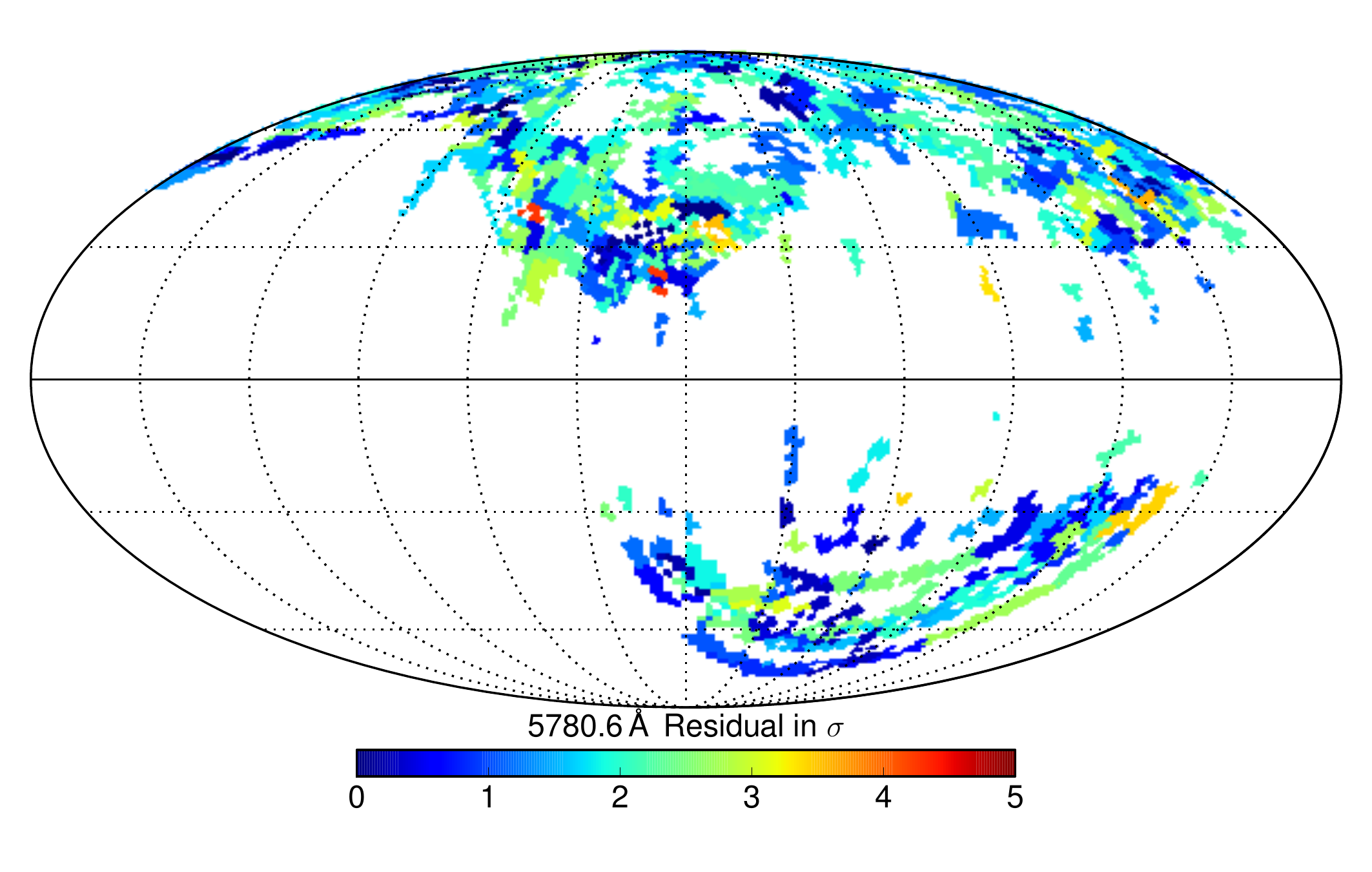}
\includegraphics[width=0.9\textwidth]{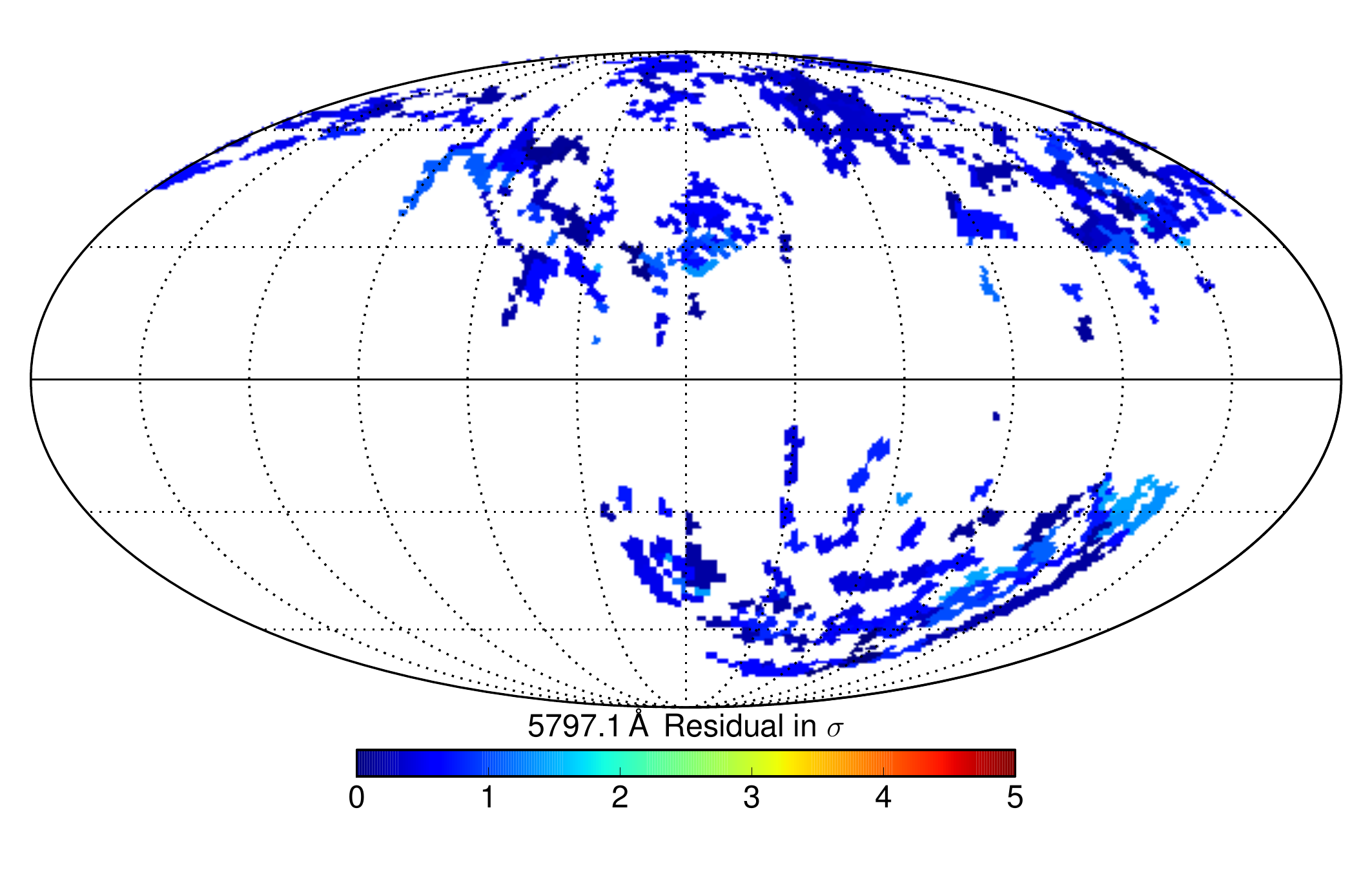}
\caption{EW residual map for the DIBs 5780.6\,\AA (top) and 5797.1\,\AA\ (bottom) normalised by the uncertainty in the residual. For a given color excess we calculate the divergence of the EW from the expected EW$_e$ -- which is based on the whole-sky determination in section \ref{s:dust} --  the normalized residual is thus  $(\mathrm{EW}-\mathrm{EW}_e)/\mathrm{EW}_e$ and its uncertainty is the quadrature of the uncertainty of EW and EW$_e$.}\label{f:res57}
\end{figure*}

Since the four lines are strongly correlated with extinction, as shown in Section \ref{s:dust}, the distribution of their EW on the sky is of little interest as it generally follows the SFD map shown in the top panel of Figure \ref{f:9}. Instead, We measure deviations from this expectation with regard to the uncertainty of the mean relation and the uncertainty of the measured EW. For every stacked spectrum, we calculate the expected equivalent width EW$_e$ based on the color-excess dependence measured over the entire sky in Section \ref{s:dust}, and show in Figure \ref{f:res57} the normalized residual, $(\mathrm{EW}-\mathrm{EW}_e)/\mathrm{EW}_e$ for every DIB. In order to present the uncertainty of the normalized residual we divide the normalized residual by its uncertainty, essentially measuring the significance of every residual value. The uncertainty of the normalized residual is calculated as the quadrature of the uncertainty of EW and the uncertainty of EW$_e$. In the absence of scatter or measurement uncertainties a residual map should be zero throughout. We see however that
the 5780.6 \AA\, map shows significant deviations from that, while the 5797.1, 6204.3 and 6613.7 \AA\, maps show deviations that are not individually significant. The high uncertainty (50 percent and more) of the mean EW-reddening relation of 5797.1, 6204.3 and 6613.7 \AA\, dominates the normalized residual uncertainty and devalues the significance of the residuals. It is therefore apparent that dust is far from being a good tracer of the 5780.6\,\AA\, DIB, as was suggested by previous studies.

\begin{figure*}
\includegraphics[width=0.9\textwidth]{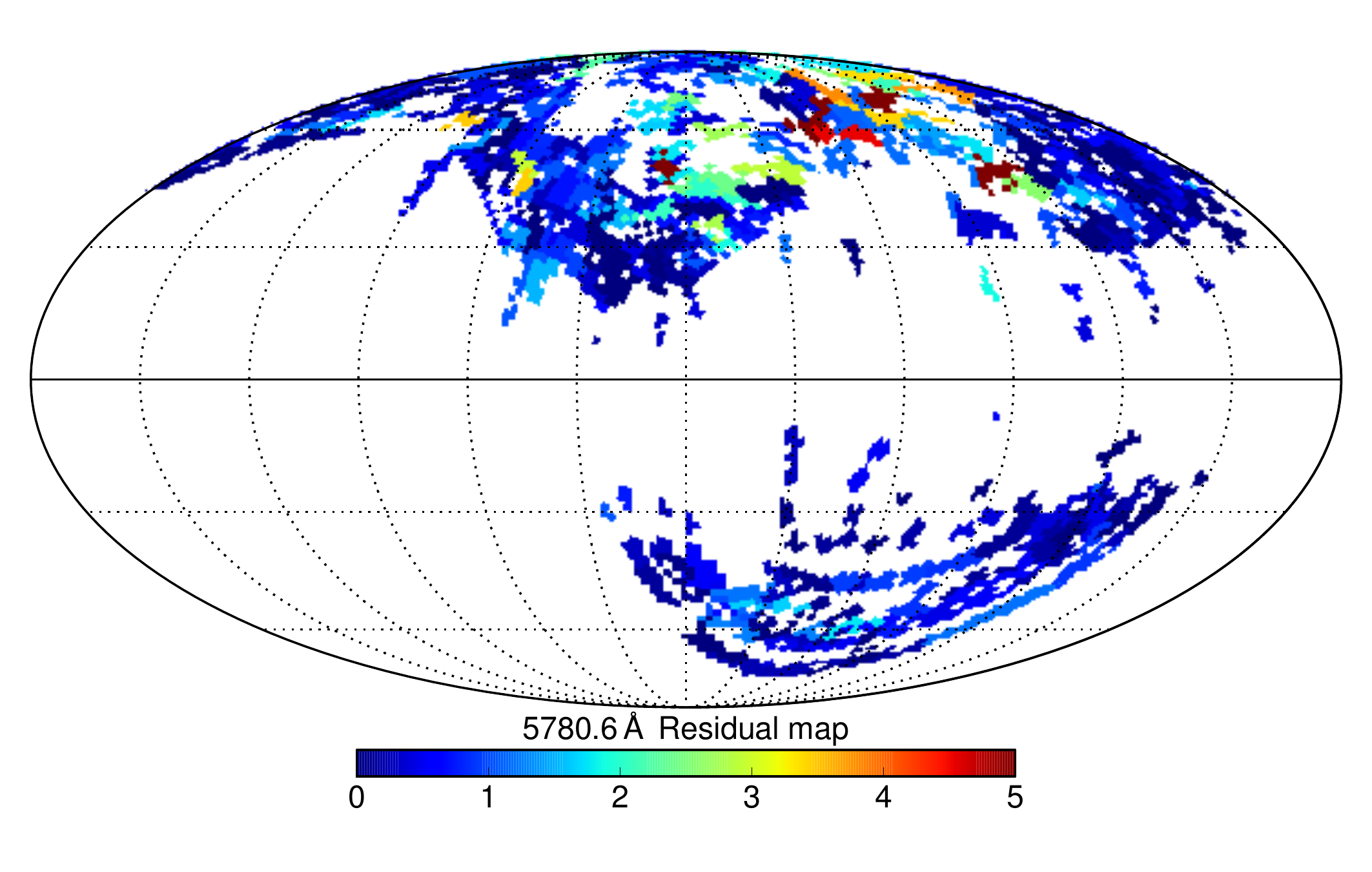}
\includegraphics[width=0.9\textwidth]{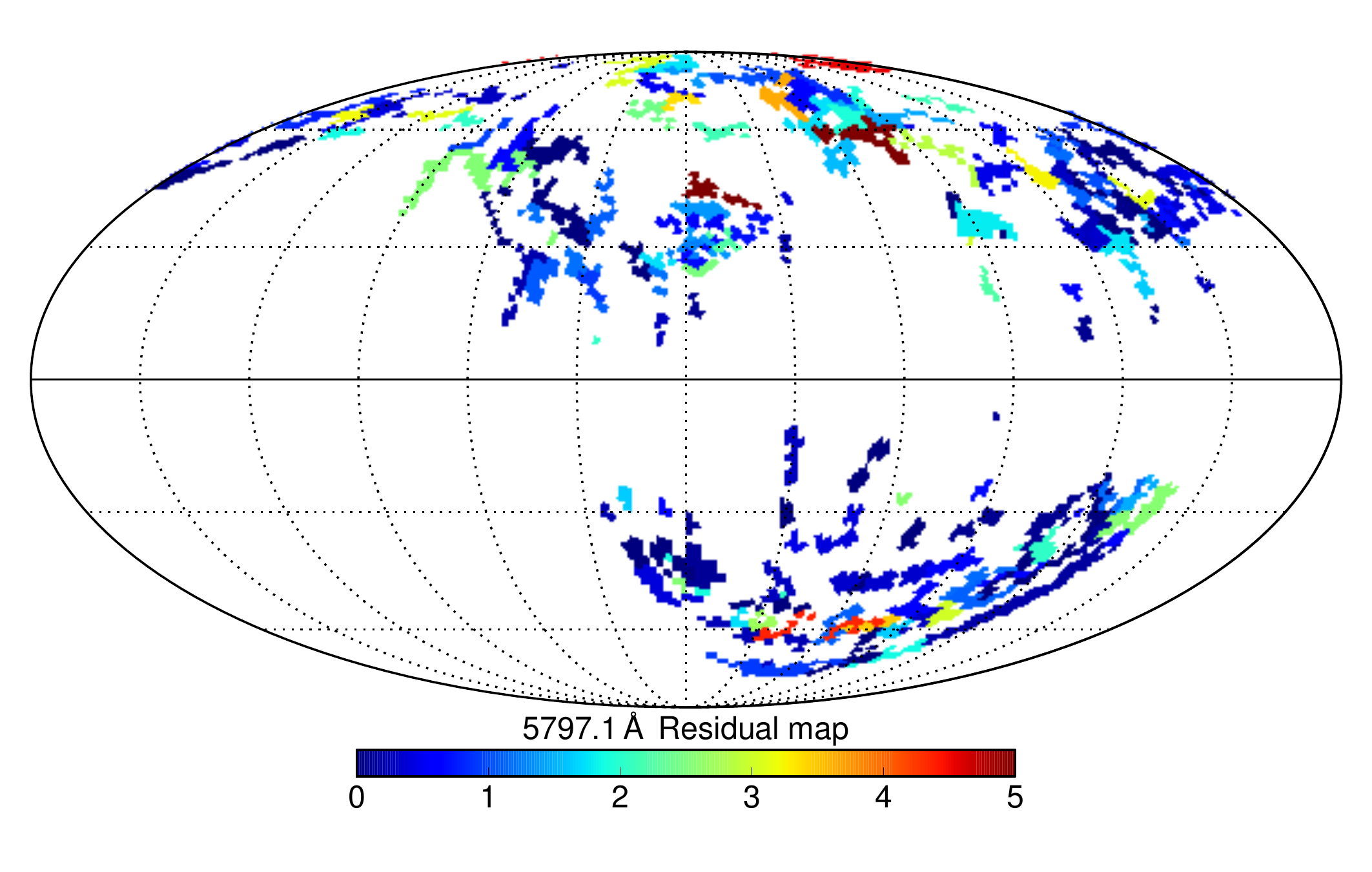}
\caption{EW residual map for the DIBs 5780.6\,\AA (top) and 5797.1\,\AA\ (bottom). For a given color excess we calculate the divergence of the EW from the expected EW$_e$ -- which is based on the whole-sky determination in section \ref{s:dust} --  the normalized residual is thus  $(\mathrm{EW}-\mathrm{EW}_e)/\mathrm{EW}_e$.}\label{f:res57_no_sig}
\end{figure*}

\begin{figure*}
\includegraphics[width=0.9\textwidth]{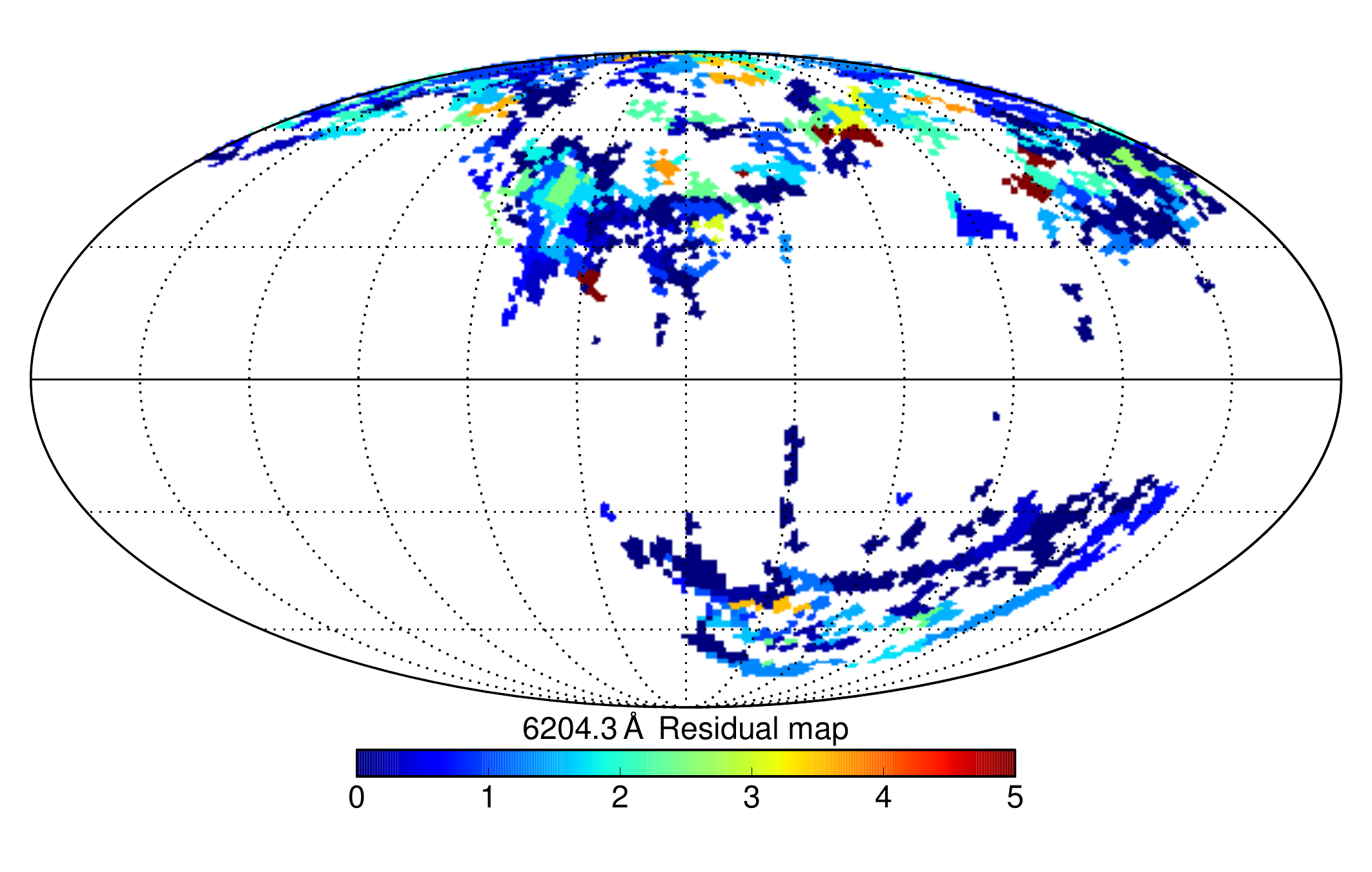}
\includegraphics[width=0.9\textwidth]{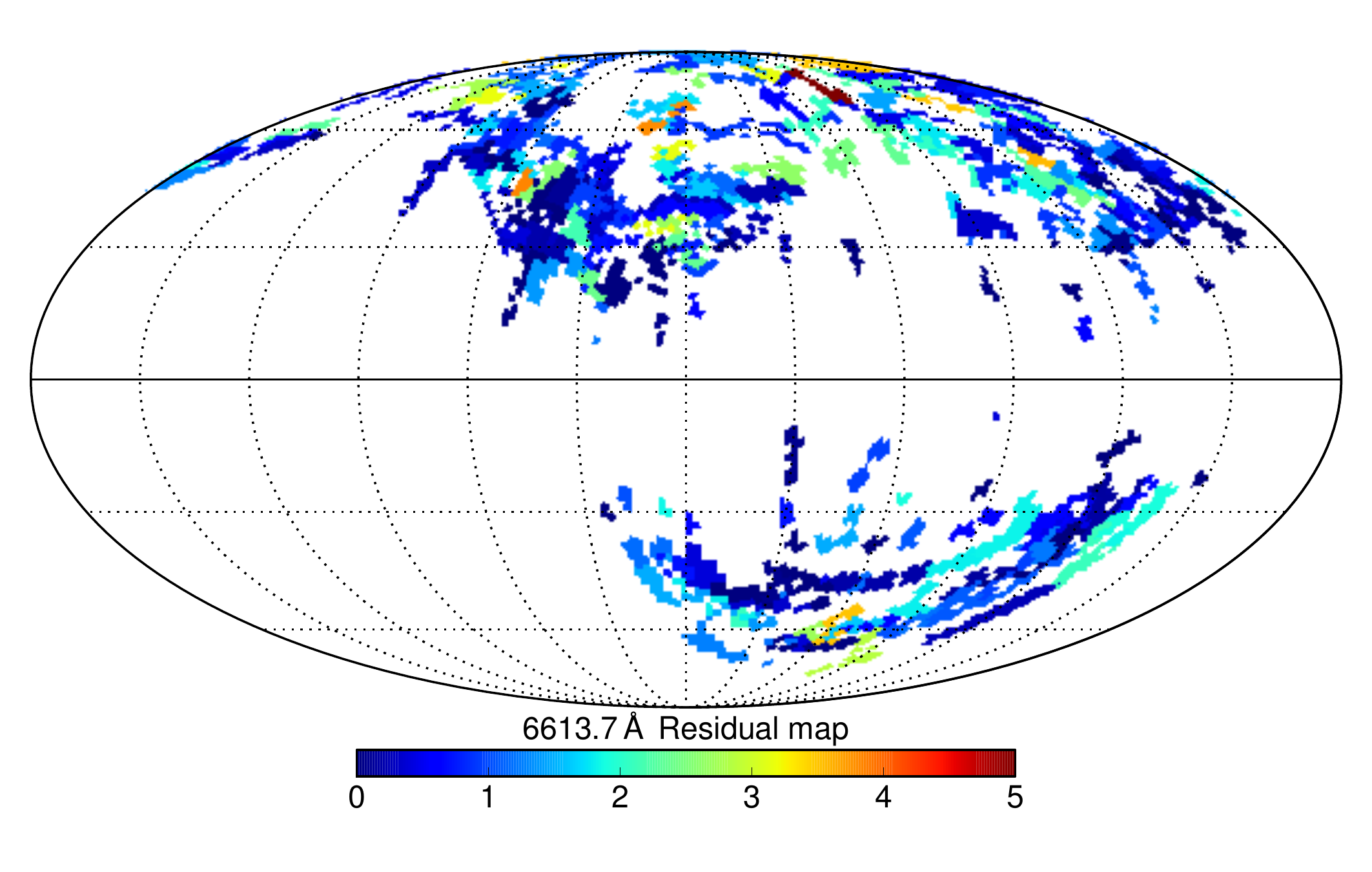}
\caption{Same as figure \ref{f:res57_no_sig} for the DIBs 6204.3\,\AA\ (top), and 6613.7\,\AA\ (bottom).}\label{f:res62_no_sig}
\end{figure*}

We present the residual maps without normalizing by the uncertainty in figures \ref{f:res57_no_sig} and \ref{f:res62_no_sig}. All the maps show deviations from the expected EW, with typically a large number of small negative residuals and a small number of high positive residuals. This imbalance is at least partially due to the fact that EWs cannot be smaller than our  7\,m\AA\ floor, but have no upper limit, so that negative divergence is capped but positive divergence is not. We note that while the residuals in a given patch of sky might be insignificant, its significance is reinforced by neighboring regions that follow the same trend.

\begin{figure*}
\includegraphics[width=3.25in]{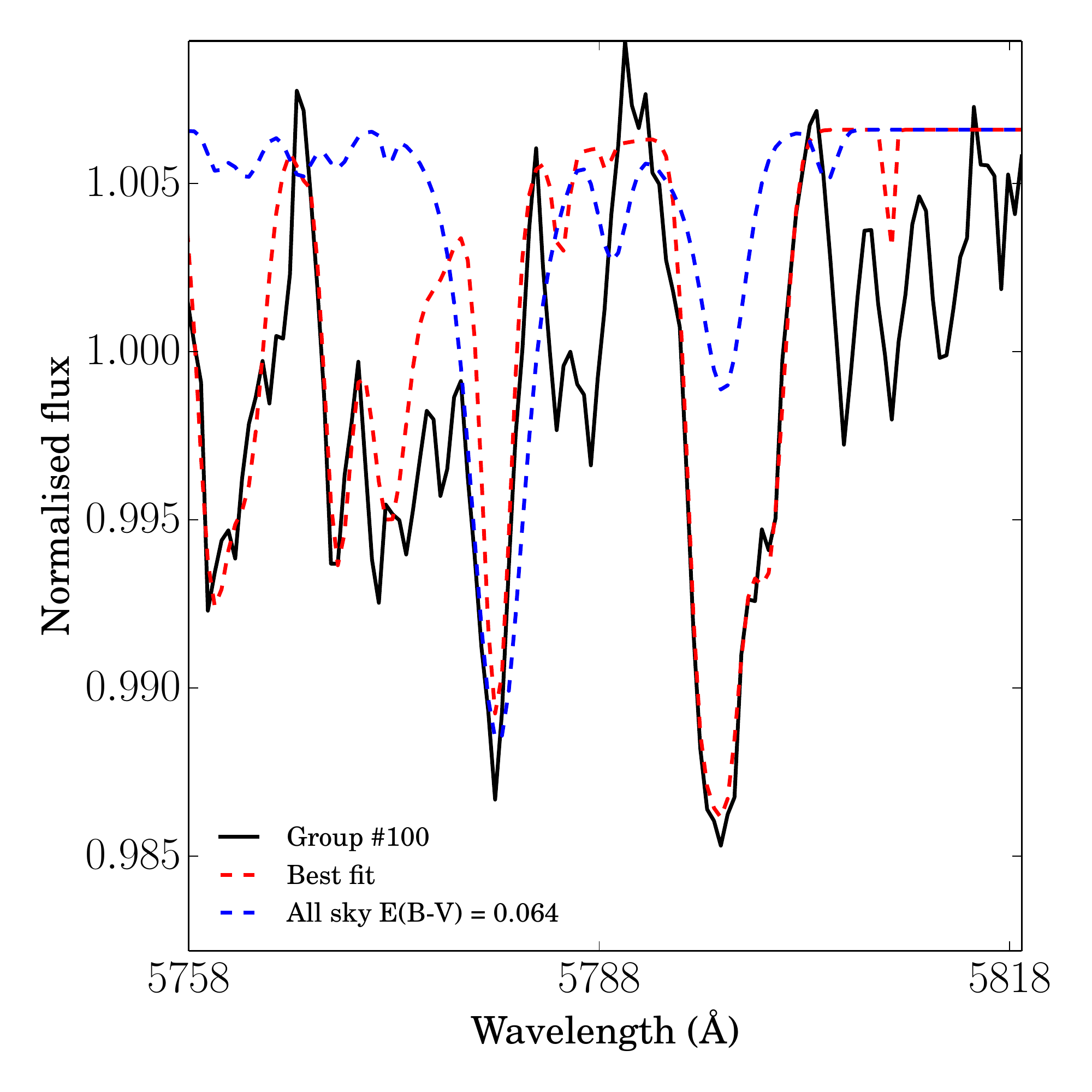}
\caption{Example spectrum (in black) of a sight line where the 5780.6\,\AA\ DIB is normal, but the 5797.1\,\AA\ is much deeper than expected when compared to sight lines with similar dust column densities (dashed blue). The best fit for the group is shown in dashed red.}\label{f:group26}
\end{figure*}

A few intriguing results are apparent when observing the residual maps. First, they are different from each other. The different DIBs are over (or under) abundant in different areas. This is an indication that that the four lines have four different carriers as was suggested by previous (e.g., \citealt{cami97}, \citealt{moutou99} and F11). Secondly, the DIB density fluctuations often cover areas of hundreds of square degrees, encompassing a few adjacent groups, as neighboring groups tend to have similar residuals. Thirdly, different groups (which consist of unrelated spectra) that diverge most strongly from the general trend and have extreme residual values (with EWs 3--5 times larger than expected) are small and isolated. In figure \ref{f:group26} we show an example for such a group -- the EW of 5780.6\,\AA\ is normal, but 5797.1\,\AA\, is much stronger than expected, and in fact stronger than 5780.6\,\AA, which is very unusual. These maps are effectively a measurement of the distribution of the various DIB carriers on the sky, and future comparisons to other ISM species may perhaps point to the carrier of each line.

\section{Conclusions}\label{s:conc}
While most of our findings have been discussed before, we have an unprecedented coverage of more than a quarter of the sky, mostly at high Galactic latitude. We study the correlations of DIBs strength with dust extinction, a handful of atomic species, and the distribution of DIBs on the sky. Our results can be summarized as such:

\begin{itemize}

\item As often determined before  (e.g, \citealt{sarre06}, F11 and K13), we find for 8 DIBs (listed in table \ref{t:1}) that their EWs correlate strongly with dust extinction, though about half of them are consistent with existing even when the dust content is negligible. 

\item The slopes for the DIB-dust relations we measure are usually steeper than previously found, but in a range of very low extinction never probed before. This might also indicate that our random high-Galactic latitude sight lines have distinct properties (e.g., density, UV radiation field, dust composition) from sight lines towards the usual DIB target-stars.

\item The slopes we find and the fact that we detect DIBs in sight lines with little to no extinction, more so at very high Galactic latitudes, may indicate that the scale height of the DIBs carriers is significant, and their distribution may be more extended than the distribution of dust (as was also recently found independently by \citealt{kos14}). 

\item We measure the relations of the 5780.6 and 5797.1\,\AA\ lines with the Na\,I\,D doublet, and the Ca\,H\&K lines. Since our data are free of stellar lines only the ISM contributes to the dependences we derive, and we find tight linear relations.  

\item We measure the distribution on the sky of the 5780.6, 5797.1 6204.3 and 6613.7\,\AA\ lines. Correcting for their correlation with dust, we find that the DIB-dust relations vary wildly on the sky (see also \citealt{vavloon09} and \citealt{van-loon13}), and we see no significant pairwise correlation between these lines, indicating that they each come from a different carrier (\citealt{cami97}, \citealt{moutou99} and F11). 

\item  From those same maps we find that the DIBs over- (and under-) densities on the sky are in large patches, up to hundreds of square degrees. Studying these patches and matching them with various ISM species may help elucidate the enduring mystery of the origin of DIBs. 

\end{itemize}

The work of \citet{lan14a} also studies the DIBs with SDSS spectra, but it is complimentary to ours. Interestingly both the method and the focus are different in the two works. We perform the source removal prior to stacking with a per-object numerical approach; they construct and subtract templates from the spectra of quasars and galaxies. They further apply their method to stellar spectra; we use only extragalactic sources. Consequently our results are mainly at very low column densities which they do not explore in depth. We study the DIBs relations with dust and metals, they study relations with dust, neutral hydrogen, molecular hydrogen, and polycyclic aromatic hydrocarbons. 

In a forthcoming publication (Baron et al. in preparation) we obviate the need  to measure EWs of lines, thus eliminating the uncertainties due to line blending and continuum determination, by studying the pairwise correlations between all DIBs. This allows us to study hundreds of DIBs, instead of the few we analyze here, and to divide the DIBs into families -- determining which other lines are associated with the stronger lines studied here. This may prove to be key predictions to laboratory experiments by determining which DIBs need to be matched in unison.

\section*{Acknowledgments}
We thank the anonymous referee, who contributed immensly to this work and its presentation.
We further thank 
Anja C. Andersen, 
D. Brailovsky,
N. Levy, 
A. Loeb,
D. Maoz,
E. Nakar, 
A. Sternberg,
and B. Trakhtenbrot 
for useful comments or valuable advice regarding different aspects of this work. 

D.P. acknowledges the support of the Alon fellowship for outstanding young researchers, and of the Raymond and Beverly Sackler Chair for young scientists. D. B. and D. P. thank the Dark Cosmology Center which is funded by the Danish National Research Foundation for hosting them while working on this topic.

The bulk of our computations was performed on the resources of the National Energy Research Scientific Computing Center, which is supported by the Office of Science of the U.S. Department of Energy under Contract No. DE-AC02-05CH11231, using the open source scientific database SciDB\footnote{www.scidb.org}. The  spectroscopic analysis was made using IPython \citep{perez07}. We also used these Python packages:  pyspeckit\footnote{www.pyspeckit.bitbucket.org}, healpy\footnote{www.healpy.readthedocs.org}, and astropy\footnote{www.astropy.org/}.

This work made extensive use of SDSS-III\footnote{www.sdss3.org} data. Funding for SDSS-III has been provided by the Alfred P. Sloan Foundation, the Participating Institutions, the National Science Foundation, and the U.S. Department of Energy Office of Science. SDSS-III is managed by the Astrophysical Research Consortium for the Participating Institutions of the SDSS-III Collaboration including the University of Arizona, the Brazilian Participation Group, Brookhaven National Laboratory, Carnegie Mellon University, University of Florida, the French Participation Group, the German Participation Group, Harvard University, the Instituto de Astrofisica de Canarias, the Michigan State/Notre Dame/JINA Participation Group, Johns Hopkins University, Lawrence Berkeley National Laboratory, Max Planck Institute for Astrophysics, Max Planck Institute for Extraterrestrial Physics, New Mexico State University, New York University, Ohio State University, Pennsylvania State University, University of Portsmouth, Princeton University, the Spanish Participation Group, University of Tokyo, University of Utah, Vanderbilt University, University of Virginia, University of Washington, and Yale University.

\bibliographystyle{mn2e}

% \section*{SUPPORTING INFORMATION}
% Additional Supporting Information may be found in the online version of this article:\\
% \item \ref{t:EW_mes_small} - EW as a function of E(B-V) for all the DIBs in the study. 
% \item Stacked spectra by color-excess.
% \item Stacked spectra by coordinates. 
% \item Sky maps that contain the following information per pixel: pixel_id, average E(B-V), 
% 
% 

\clearpage

\end{document}